\documentclass[prl,aps,superscriptaddress,footinbib,twocolumn]{revtex4-2}
\usepackage[latin9]{inputenc}
\usepackage{color}
\usepackage{amsmath}
\usepackage{amssymb}
\usepackage{stmaryrd}
\usepackage{graphicx}
\usepackage[normalem]{ulem}
\usepackage{bm}
\usepackage[ruled]{algorithm2e}
\usepackage[unicode=true,bookmarks=true,bookmarksnumbered=false,bookmarksopen=false,breaklinks=false,pdfborder={0 0 1},backref=false,colorlinks=true]
 {hyperref}
\hypersetup{
 linkcolor=magenta, urlcolor=blue, citecolor=blue, pdfstartview={FitH}, hyperfootnotes=false, unicode=true}
 
\usepackage{todonotes}
\makeatletter

\usepackage{amsfonts}
\usepackage{dcolumn}
\usepackage{graphicx}
\usepackage{epstopdf}
\usepackage{times}
\usepackage{xcolor}
\usepackage{soul}
\def\ket#1{\left|#1\right\rangle}
\def\bra#1{\left\langle#1\right|}

\makeatother

\begin{document}

\title{Experimental demonstration of reconstructing quantum states with generative models}

\author{Xuegang Li}
\thanks{These authors contributed equally to this work.}
\affiliation{Center for Quantum Information, Institute for Interdisciplinary Information Sciences, Tsinghua University, Beijing 100084, China}

\author{Wenjie Jiang}
\thanks{These authors contributed equally to this work.}
\affiliation{Center for Quantum Information, Institute for Interdisciplinary Information Sciences, Tsinghua University, Beijing 100084, China}

\author{Ziyue Hua}
\thanks{These authors contributed equally to this work.}
\affiliation{Center for Quantum Information, Institute for Interdisciplinary Information Sciences, Tsinghua University, Beijing 100084, China}

\author{Weiting Wang}
\email{wangwt2020@mail.tsinghua.edu.cn}
\affiliation{Center for Quantum Information, Institute for Interdisciplinary Information Sciences, Tsinghua University, Beijing 100084, China}

\author{Xiaoxuan Pan}
\affiliation{Center for Quantum Information, Institute for Interdisciplinary Information Sciences, Tsinghua University, Beijing 100084, China}

\author{Weizhou Cai}
\affiliation{Center for Quantum Information, Institute for Interdisciplinary Information Sciences, Tsinghua University, Beijing 100084, China}

\author{Zhide~Lu}
\affiliation{Center for Quantum Information, Institute for Interdisciplinary Information Sciences, Tsinghua University, Beijing 100084, China}

\author{Jiaxiu Han}
\affiliation{Center for Quantum Information, Institute for Interdisciplinary Information Sciences, Tsinghua University, Beijing 100084, China}

\author{Rebing Wu}
\affiliation{Center for Intelligent and Networked Systems,
Department of Automation, Tsinghua University, Beijing 100084, P. R. China}
\affiliation{Hefei National Laboratory, Hefei 230088, China}

\author{Chang-Ling Zou}
\affiliation{Key Laboratory of Quantum Information, CAS, University of Science and Technology of China, Hefei, Anhui 230026, P. R. China}
\affiliation{Hefei National Laboratory, Hefei 230088, China}

\author{Dong-Ling Deng}\email{dldeng@tsinghua.edu.cn}
\affiliation{Center for Quantum Information, Institute for Interdisciplinary Information Sciences, Tsinghua University, Beijing 100084, China}
\affiliation{Shanghai Qi Zhi Institute, 41th Floor, AI Tower, No. 701 Yunjin Road, Xuhui District, Shanghai 200232, China}
\affiliation{Hefei National Laboratory, Hefei 230088, China}

\author{Luyan Sun}\email{luyansun@tsinghua.edu.cn}
\affiliation{Center for Quantum Information, Institute for Interdisciplinary Information Sciences, Tsinghua University, Beijing 100084, China}
\affiliation{Hefei National Laboratory, Hefei 230088, China}

\begin{abstract} 
  Quantum state tomography, a process that reconstructs a quantum state from measurements on an ensemble of identically prepared copies,  plays a crucial role in benchmarking quantum devices. However, brute-force approaches to quantum state tomography would become impractical for large systems, as the required resources scale exponentially with the system size. Here, we explore a machine learning approach and report an experimental demonstration of reconstructing quantum states based on neural network generative models with an array of programmable superconducting transmon qubits. In particular, we experimentally prepare the Greenberger-Horne-Zeilinger states and random states up to five qubits and demonstrate that the machine learning approach can efficiently reconstruct these states with the number of required experimental samples scaling linearly with system size. Our results experimentally showcase the intriguing potential for exploiting machine learning techniques in validating and characterizing complex quantum devices, offering a valuable guide for the future development of quantum technologies. 
  \end{abstract}
  
  \maketitle

With the rapid development of quantum devices across various platforms~\cite{Harrigan2021Quantum,Kandala2017Hardwareefficient,Arute2019Quantum,Song2019Generation,Yan2019Strongly,Ma2019Dissipatively,Kollar2019Hyperbolic,Andersen2020Repeated,Jurcevic2021Demonstration,Arute2019Quantum,GOOGLEAIQUANTUMANDCOLLABORATORS2020HartreeFock,Bluvstein2024Logical,Song2019Generation}, reconstructing quantum many-body states from experimentally measured data posts a crucial challenge. Straightforward quantum state tomography (QST) is only applicable for small systems~\cite{Hawkes2003Advances,Cramer2010Efficient}, since the required classical computing resources, such as the number of measurements and the memory size, grow exponentially as the system size increases. Such exponential difficulty comes from the exponential growth of the dimensionality of the corresponding Hilbert space, rendering this method inefficient for reconstructing large quantum many-body systems.

  From a more physical perspective, the quantum states  encountered in experiments are not randomly chosen from the entire Hilbert space, but rather from a more structured and confined region within that space~\cite{Heinosaari2013Quantum,Cramer2010Efficient}.
  Thus, in various physical scenarios, prior information about the target quantum state will be encoded into the structure of a parameterized ansatz, and the variational parameters in such an ansatz will be optimized to approximate the distribution of the experimental measurement data. Many quantum tomography methods have been proposed to address the challenge of reconstructing quantum states in this direction~\cite{Orus2014Practical,Schollwock2011Densitymatrix,Carleo2017Solving,Deng2017Quantum,Torlai2018Neuralnetwork,Gao2017Efficient,Glasser2018NeuralNetwork}. 
  Quantum tomography using matrix product states~\cite{Orus2014Practical,Schollwock2011Densitymatrix} is a favorable choice when the target state comes from a one-dimensional chain~\cite{Lanyon2017Efficient}. Nevertheless, this ansatz poses constraints for the system's dimensionality and entanglement entropy, which limits its applications.
  Restricted Boltzmann machine~\cite{Carleo2017Solving,Deng2017Quantum,Torlai2018Neuralnetwork,Gao2017Efficient,Glasser2018NeuralNetwork} features non-local connections and can be applied to reconstruct high dimensional and extensively entangled quantum states~\cite{Carleo2017Solving,Deng2017Machine,Deng2017Quantum,Torlai2018Neuralnetwork}. However, its extension to to mixed states through purification is also resource demanding in general~\cite{Torlai2018Latent}, which limits its applicability in the noisy intermediate-scale quantum (NISQ) era~\cite{Preskill2018Quantum}. In Ref.~\cite{Carrasquilla2019Reconstructing}, a generative model based on recurrent neural networks (RNNs)~\cite{Pascanu2013Difficulty,Salehinejad2018Recent,Fuchs2017SIC,Sutskever2011Generating} is employed to describe noisy quantum many-body states classically. 

  \begin{figure*}[t]
    \includegraphics[width=0.8\textwidth]{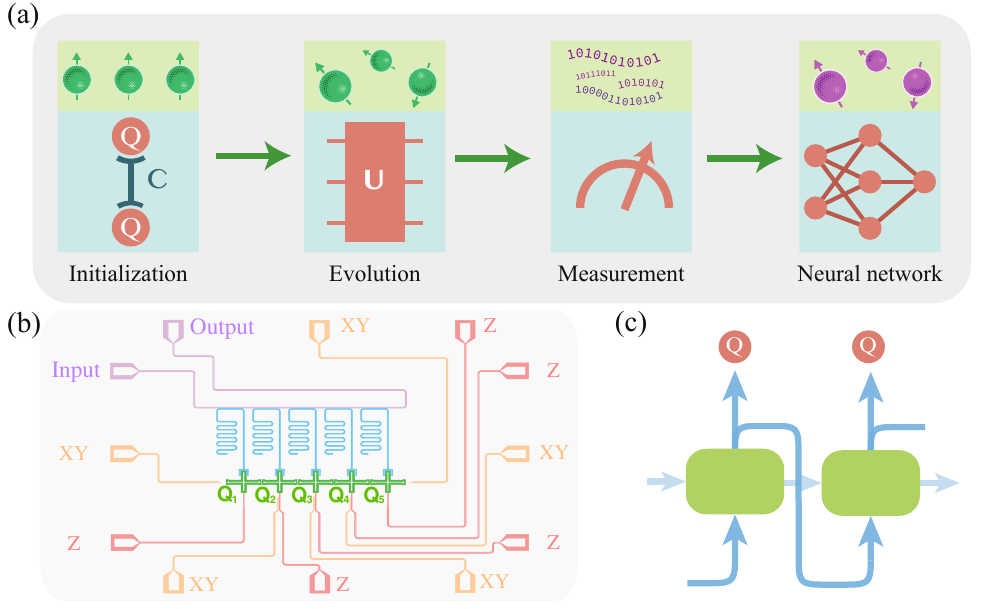}
    \caption{ {Schematic of the quantum state reconstruction with a recurrent neural network (RNN).} (a) Schematic of the experiment. All qubits are initialized at the ground states and a quantum circuit is applied to them to prepare a many-body state. Following, the resulting quantum states are measured by a randomly chosen projector. The state preparation and measurement are repeated many times, and the projective measurement outcomes are used to train the RNN-based generative model. When the training process is finished, the distribution generated by the RNN can be used to reconstruct the state of the system or to calculate physical quantities. (b) The five-transmon-qubit sample is used in this work. All the qubits are connected as a one-dimensional chain. Each qubit can be read out and controlled individually in the $Z$ and $XY$ directions, respectively. (c) Schematic of the RNN model, where the hidden state (light blue arrows between green blocks) passes information through different sites.}
    \label{figs:Schematics}
  \end{figure*}
  
 In this paper, we report an experimental demonstration of the generative-model-based tomography method with an array of programmable superconducting transmon qubits. We first prepare the Greenberger-Horne-Zeilinger (GHZ) state for $N = 2,\, 3,\, 4,\, 5$ qubits on our superconducting processor with non-negligible errors. The corresponding measured QST fidelities are $98.0\%, 97.9\%, 93.3\%, 89.4\%$, respectively. 
  Then, we reconstruct these noisy GHZ states using the RNN-based generative model with prior hyper-parameters. We observe that this learning method manifests a fast convergence rate and the reconstructed states from the trained model agree well with the experimental states.
  We show that the required classical computational resource scales linearly with the system size. 
  In addition, we demonstrate the ability to extract useful information from the classical description of the quantum states and compute the $\langle XX\rangle$ and $\langle ZZ\rangle$ correlations using the well-trained generative model, where the results closely align with the experimental measurements. Our results demonstrate the potential of neural-network-based tomography and manifest the advantages of using the classical generative model to reconstruct complex noisy quantum states.

We start with a fixed initial state and apply quantum circuits to prepare our target state. It has been shown that a generative model based on an RNN can be used to reconstruct many physical states, ranging from the extensively correlated cat state to the ground state of a many-body system~\cite{Carrasquilla2019Reconstructing}. Here, we apply such a model to reconstruct GHZ states.  
  GHZ states play a central role in many quantum information applications, including secret sharing~\cite{Hillery1999Quantum} and quantum sensing~\cite{Degen2017Quantum}. Despite their applications, GHZ states are also cat states characterized by macroscopic correlations and are fragile to many practical noises. In our experiment, we prepare GHZ states with different system sizes with our superconducting processor. Subsequently, we carry out a positive operator-valued measure (POVM), which is informationally complete~\cite{Fuchs2017SIC,Scott2006Tight,DAriano2004Informationally}, and collect the measurement results. After repeating these procedures many times, we send the collected measurement results into an RNN-based generative model and optimize its variational parameters to learn the distribution of the measurements [see Fig.~\ref{figs:Schematics}(a) for a pictorial illustration]. 
 The resulting measurement distribution can then be used to determine the quantum states.
    
  \begin{figure*}[t]
    \includegraphics[width=0.96\textwidth]{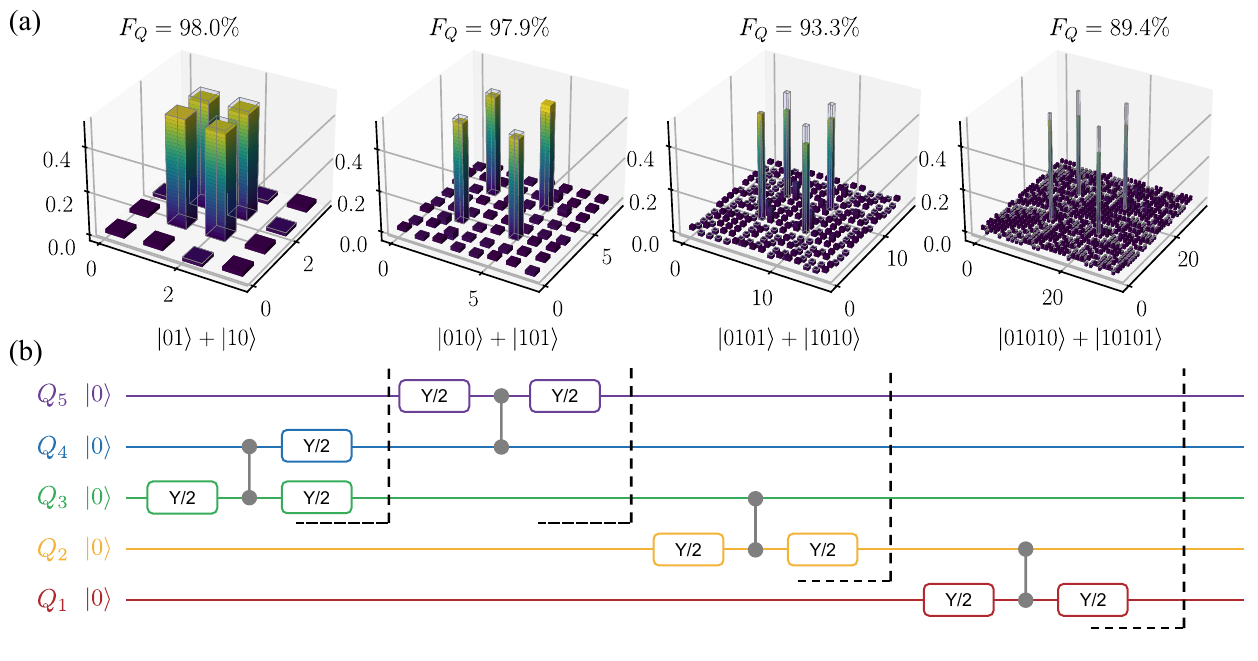}
    \caption{ {Experimental quantum state tomography of the GHZ states.} (a) The prepared two-qubit Bell state and three-, four-, and five-qubit GHZ states in the experiment. Only the real parts of the corresponding density matrices $\rho$ are shown. All the imaginary parts are nearly zero and are ignored in this figure. The transparent bars represent the real part of the ideal density matrix $\rho_\mathrm{ideal}$. The decimal numbers on the $X$ and $Y$ axes represent the binary number states. For example, the number $4$ represents the $\ket{00100}$ state in a five-qubit system. (b) Quantum circuits for preparing the five-qubit GHZ state. The two-qubit Bell state is prepared on $Q_3$ and $Q_4$. The three-, four-, and five-qubit GHZ states are prepared following the sequence up to the respective black dashed lines.} 
    \label{figs:ExpState}
  \end{figure*}
  
  In our experiment, we demonstrate the above quantum state reconstruction strategy to recover practical quantum states prepared in a superconducting circuit, where five cross-shaped transmon superconducting qubits~\cite{Barends2013Coherent,Barends2014Superconducting} are connected in a linear array as shown in Fig.~\ref{figs:Schematics}(b). 
  Each qubit has independent $X$, $Y$, and $Z$ controls and can be read out individually. The measurement of the qubits is based on the dispersive interaction between the readout resonator and the corresponding qubit~\cite{Schuster2007Resolving}. We use a Josephson parametric amplifier~\cite{Kamal2009Signaltopump,Vijay2009Invited,Roy2015Broadband} with a bandwidth of 260~MHz and a gain of more than 20~dB to facilitate the readout of all qubits simultaneously. More device parameters can be found in Refs.~\cite{Li2018Perfect,supplement}.
  
 We mainly consider the two-qubit Bell state and the multi-qubit GHZ states. We prepare these states with single-qubit gates and two-qubit controlled-phase (CZ) gates, as shown in Fig.~\ref{figs:ExpState} (see Sec.~II.B in Ref.~\cite{supplement} for more details). The two-qubit Bell state is prepared on $Q_3$ and $Q_4$, which operate close to their sweet spots, thus exhibiting the highest coherence times and higher fidelity. To reduce the effect of the residual $ZZ$ coupling between neighboring qubits, we prepare the Bell state $\ket{01}+\ket{10}$ instead of $\ket{00}+\ket{11}$, as it is immune to the entanglement phase accumulation. We prepare the three-, four-, and five-qubit GHZ states following the same strategy. We note that these states can be transformed to the original Bell (GHZ) state using a single layer of $X$ gates, regardless of the number of qubits.
  
  We then measure the prepared quantum states. 
  The single-qubit state can be characterized with the Pauli-4 POVM \cite{Carrasquilla2019Reconstructing}, defined as $M_{\mathrm{Pauli}-4}=\{M^{(0)}=\frac{1}{3}\ket{0}\bra{0},~M^{(1)}=\frac{1}{3}\ket{+}\bra{+},~M^{(2)}=\frac{1}{3}\ket{l}\bra{l},~M^{(3)}=\frac{1}{3}\{\ket{1}\bra{1}+\ket{-}\bra{-}+\ket{r}\bra{r}\}\}$, where \{$\ket{+}$,~$\ket{-}$\} and \{$\ket{r}$,~$\ket{l}$\} are eigenstates of Pauli operators $\sigma_x$ and $\sigma_y$, respectively. 
  The projection measurement of each qubit along different axes on the Bloch sphere can be realized by applying a suitable pre-rotation gate followed by a $\sigma_Z$ measurement. The joint projection measurement on multiple qubits can be achieved by performing individual projections on each qubit.
  To reconstruct the quantum state, we randomly choose one of the projectors ($\sigma_x$, $\sigma_y$, or $\sigma_Z$) to measure the prepared quantum state and record the measurement results. We repeat this randomized measurement many times, and use the collected results to carry out the standard QST or the tomography based on the generative model.

  We note that the measured raw data includes all kinds of errors, such as the readout imperfections. Consequently, after finishing the learning process the learned quantum state also suffers from these errors. To obtain information of interest more accurately from this model, such as the Pauli correlations between qubits, several improvements can be performed. For example, Bayes' rule can be applied to infer a more accurate measurement probability distribution and to alleviate the readout imperfection. Besides, maximum-likelihood estimation (MLE) used in standard QST can also be utilized when we reconstruct the density matrix from the trained generative model to ensure that the corresponding density matrix is physically meaningful, i.e., it is a hermitian and positive semi-definite matrix with trace equal to one.
  We note that both the Bayes correction and the MLE method require exponential resources. However, when the quality of both quantum devices and quantum operations is further improved in the future, we can directly use the raw data without Bayes correction to train our generative model. As for the MLE, we remark that the density matrix is not necessary and the classical distribution is sufficient in many scenarios, where the MLE method is avoided.
    
We  benchmark the prepared states by the standard QST where the fidelities of the Bell (GHZ) state for different system sizes are $\mathrm{Tr}(\sqrt{\rho^{1/2}\rho_\mathrm{{ideal}}\rho^{1/2}})=98.0\%\ (N=2), 97.9\%\ (N=3), 93.3\%\ (N=4)$, and $89.4\%\ (N=5)$. The extracted density matrices $\rho$ of the Bell state and the GHZ states are shown in Fig.~\ref{figs:ExpState}(a). The real parts of the experimental density matrices $\rho$ are shown in the top panel, while the imaginary parts are almost zero and not shown. The transparent bars represent the real parts of the ideal density matrices $\rho_\mathrm{{ideal}}$. 
  
  \begin{figure}[t]
    \centering
    \includegraphics[width=0.48\textwidth]{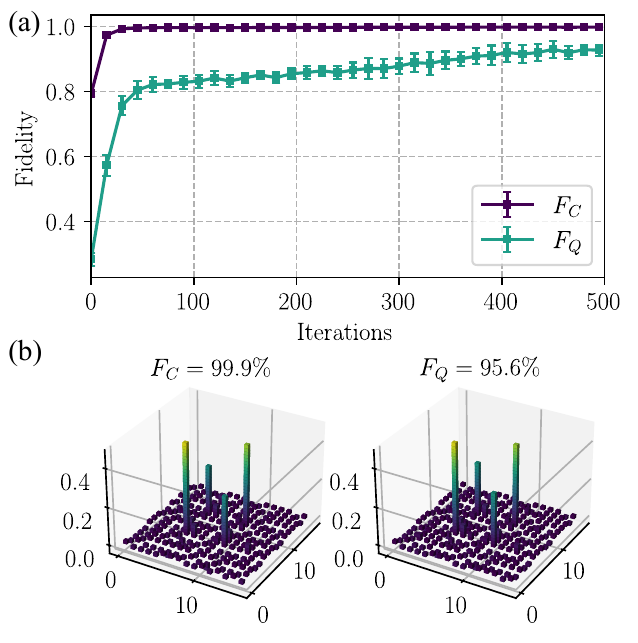}
    \caption{ {Reconstructed four-qubit GHZ state with RNN.} (a) The training process of four-qubit GHZ state. The dark purple and cyan lines represent the classical and quantum fidelities, respectively.
    (b) The real parts of the reconstructed four-qubit GHZ state with the maximum classical fidelity and with the maximum quantum fidelity, respectively.}
    \label{figs:ExpLearning}
  \end{figure}
  
  We then explore the performance of the quantum tomography method based on the generative model. Particularly, we focus on the GHZ state with $N=4$. 
  The classical description learned by the RNN model is optimized by minimizing the KL-divergence $\sum_i p_i\log\tilde{p}_i$ between the experimentally measured distribution  $P_\mathrm{{exp}}$ and the generated distribution $P_\mathrm{{model}}$ by the RNN. The classical fidelity $F_\mathrm{C}=\sum_{P_\mathrm{{exp}}}(P_\mathrm{{model}}/P_\mathrm{{exp}})^{1/2}$ and quantum fidelity $F_\mathrm{Q}=\mathrm{Tr}(\sqrt{\rho_\mathrm{{model}}^{1/2}\rho_\mathrm{{exp}}\rho_\mathrm{{model}}^{1/2}})$ are calculated from the learned Pauli-4 POVM distributions [see Fig.~\ref{figs:ExpLearning}(a)]. 
  We observe that the learning process converges quickly. 
  The maximum classical fidelity $F_\mathrm{C}$ is higher than $99\%$ while the quantum fidelity $F_\mathrm{Q}$ is near $95\%$, indicating the quantum fidelity is more sensitive to the learning errors. Nevertheless, we notice that both the classical and the quantum fidelities gradually approach the maximum values with a similar learning behavior.
  To demonstrate the learning result, we depict the density matrices of the learned states with the maximum classical fidelity and the maximum quantum fidelity in Fig.~\ref{figs:ExpLearning}(b), respectively. These two states exhibit a strong resemblance, as reflected by their similar quantum fidelities compared to the reconstructed four-qubit GHZ state using the QST method ($F_\mathrm{Q}=\ 94.6\%$ vs $95.6\%$).
  Therefore, $F_\mathrm{C}$ can be regarded as a benchmark when $F_\mathrm{Q}$ is in practice difficult to calculate for a large system size $N$.
  
  \begin{figure}[t]
    \centering
    \includegraphics[width=0.48\textwidth]{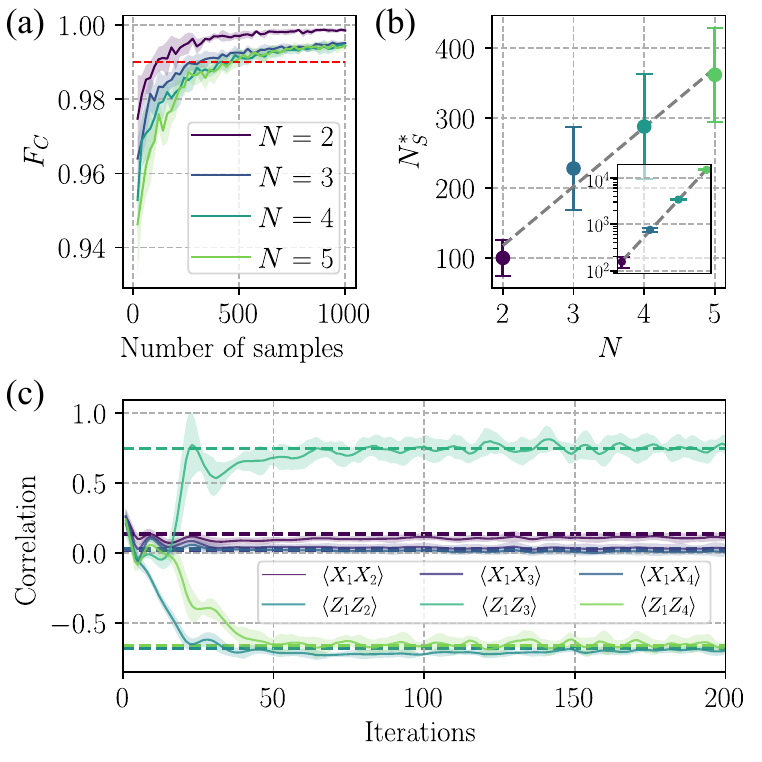}
    \caption{ {Scaling of the number of samples for the generative-model-based quantum tomography.} (a) The average classical fidelities as a function of the necessary number of samples $N_\mathrm{s}$ performed on the GHZ state with $N = 2,\,3,\,4,$ and $5$ qubits, respectively. The black dashed line denotes the reference fidelity of $99\%$. The error bars are obtained by repeating the tomography process $10$ times by randomly choosing $N_\mathrm{s}$ samples from the experimental data. (b) Scaling of the necessary $N_\mathrm{s}$ for reconstructing the GHZ states using the RNN model. The deep orange dashed line denotes a linear fit, indicating the powerful representation of the RNN model. 
    Inset: Necessary number of samples for reconstructing GHZ states using the QST method, which shows a clear exponential scaling. (c) Reconstructed two-body $XX$ and $ZZ$ correlations after the maximum-likelihood-estimation correction. The corresponding experimental correlations computed using the QST method are shown as the dashed lines.}
    \label{figs:ExpScaling}
  \end{figure}

  Furthermore, we investigate the resources required for the generative-model-based quantum tomography. 
  We calculate the average classical fidelity between the distribution generated by the trained RNN and the distribution of the total experimental measurements, treating it as a function of the training set size $N_\mathrm{s}$. 
  For each total number of qubits $N$, we repeat the learning process $10$ times by randomly choosing $N_\mathrm{s}$ samples from the experimental data. The average classical fidelities and the associated error are calculated for the $N$-qubit GHZ states and shown in Fig.~\ref{figs:ExpScaling}(a). 
  We denote the critical value $N_\mathrm{s}^*$ as the minimum number of samples from which the reconstructed GHZ (Bell) state manifests a classical fidelity larger than the reference fidelity of $99\%$.
  We plot the critical value $N_\mathrm{s}^*$ as a function of the total qubit number $N$, as shown in Fig.~\ref{figs:ExpScaling}(b).
  The gray line denotes a linear fit, indicating an approximately linear growth of the number of measurements required by the generative-model-based tomography. 
  As a comparison, the necessary number of samples for reconstructing the quantum states based on the QST method shows an exponential scaling with the number of qubits, as shown in the inset of Fig.~\ref{figs:ExpScaling}(b).
  Such a significant difference demonstrates that the generative-model-based quantum tomography method is more scalable for noisy states in real experiments.

  
  
  Finally, we demonstrate the ability to extract physical information from the quantum state represented by the well-trained RNN model. 
  For instance, we compute the $XX$ and $ZZ$ correlations between different pairs of qubits. Despite the initial deviation, all $ZZ$ correlations quickly converge to the experimentally measured ones, which are denoted as the dashed lines in Fig.~\ref{figs:ExpScaling}(c). The $XX$ correlations in the GHZ states are expected to be zero. Nevertheless, due to experimental noise, they do not exactly reach zero. Our generative model can partially capture these errors, even though the underlying errors are not fully considered when we choose the neural network ansatz.
  
  The generative-model-based quantum tomography method is not limited to highly entangled GHZ states; it can also be applied to a diverse set of experimental noisy quantum states, including general random quantum states prepared without prior knowledge (see Sec.~II.C in Ref.~\cite{supplement}).
  We observe that the reconstructed results for randomly selected quantum states behave similarly to those for the GHZ states and the required number of measurements shows an approximately linear scaling, underscoring the potential of the generative-model-based tomography method.
  
In summary, we have demonstrated experimentally the generative-model-based quantum tomography method for reconstructing noisy quantum states with a superconducting processor.
  We first measure such noisy quantum states using an informationally complete POVM, followed by utilizing the collected samples to train an RNN model with moderate variational parameters. The RNN-based generative model exhibits good performance in learning the measurement distribution, with a fast convergence rate during the learning procedure. 
  We find the necessary experimental samples  to achieve a classical fidelity of $99\%$ scale approximately linearly with the total qubit number. 
  In addition, We compute the $XX$ and $ZZ$ correlations using the trained RNN-based generative model, showcasing excellent alignment with the experimentally measured values. 
  Our work demonstrates that advanced machine learning algorithms in conjunction with quantum information tools, such as informationally complete POVMs, can be a powerful approach for characterizing quantum devices in the NISQ era.
  
\vspace{.5cm}
\begin{acknowledgments}
We would like to thank Juan Carrasquilla, Sirui Lu, Pei-Xin Shen, and Weikang Li for useful discussions.
This work is supported by National Key Research and Development Program of China under Grant No. 2016YFA0301902 and 2017YFA0304303, and National Natural Science Foundation of China (Grant Nos. 92165209, 11925404, 92365301, 11874235, 11674060, T2225008, and 12075128), the Innovation Program for Quantum Science and Technology (Grant Nos. 2021ZD0300203, 2021ZD0302203), Tsinghua University Dushi Program, and Shanghai Qi Zhi Institute.
\end{acknowledgments}

%

\cleardoublepage

\onecolumngrid
\renewcommand{\thefigure}{S\arabic{figure}}
 \setcounter{figure}{0}
\global\long\def\thepage{S\arabic{page}}%
 \setcounter{page}{1}
 \global\long\def\theequation{S\arabic{equation}}%
 \setcounter{equation}{0}
\global\long\def\tablename{\textbf{Supplementary Table}}%

\begin{center}
\textbf{\Large{}Supplementary Information for ``Experimental demonstration of reconstructing quantum states with generative models''}{\Large\par}
\par\end{center}

\section{GENERATIVE-MODEL-BASED QUANTUM TOMOGRAPHY}
\subsection{An overview of the reconstruction method}
We use generative models based on recurrent neural networks (RNNs) to reconstruct quantum states, which are prepared in a superconducting processor and are measured using a set of informationally complete projective measurements. 
This process consists of two parts: the state preparation and the state reconstruction [Fig.~1(a) of the main text]. 

As mentioned in our main text, we first prepare multiple-qubit Greenberger-Horne-Zeilinger (GHZ) states using superconducting transmon qubits, and then measure each qubit with measurement operators randomly chosen from an informationally complete positive operator-valued measure (POVM). Specifically, we focus on Pauli-4 POVM measurements in the main text. For a single qubit, the POVM is defined as 
\begin{equation}
  M_{\text {Pauli-4 }}=\left\{M^{(0)}=\frac{1}{3}|0\rangle\langle 0|,\ M^{(1)}=\frac{1}{3}|+\rangle\langle+|, \ M^{(2)}=\frac{1}{3}| l\rangle\langle l|, \ M^{(3)}=\frac{1}{3}\left(|1\rangle\langle 1|+|-\rangle\langle-|+|{r}\rangle\langle{r}|\right)\right\},
\label{eq:Pauli4}
\end{equation}
where \{$\ket{+},\ \ket{-}$\}, \{$\ket{l},\ \ket{r}$\}, and \{$\ket{0},\ \ket{1}$\} are the eigenstates of Pauli operators $\sigma_x$, $\sigma_y$, and $\sigma_z$, respectively. 
For $N$ qubits, the measurements are obtained by the tensor product of the single qubit POVM elements: 
\begin{equation}
  {M}^{(\bm{a})}={M}^{(a_1)}\otimes{M}^{(a_2)}\otimes\cdots\otimes{M}^{(a_N)},
\end{equation}
where $a_k=0,1,2,3$.
We note that the Pauli-4 POVM is informationally complete, meaning that the measurement distribution $P=\left\{P(\bm{a})=\text{Tr}\left(M^{(\bm{a})}\rho\right) \text{for all possible measurement operators } M^{(\bm{a})}\right\}$ contains all the information of the state $\rho$. As a result, such a classical distribution can be utilized to reconstruct the corresponding density matrix:
\begin{equation}
    \rho=\sum_{\bm{a}, \bm{a}^{\prime}} P(\bm{a}) T_{\bm{a}, \bm{a}^{\prime}}^{-1} M^{(\bm{a}^{\prime})}=\mathbb{E}_{\bm{a} \sim {P}}\left(\sum_{\bm{a}^{\prime}} T_{\bm{a}, \bm{a}^{\prime}}^{-1} M^{(\bm{a})}\right),\label{equ:staterepresentation}
\end{equation}
where $T_{\bm{a}, \bm{a}^{\prime}}=\text{Tr}\left(M^{(\bm{a})}M^{(\bm{a}^\prime)}\right)$ is the overlap matrix for the predefined informationally complete POVM.

Using Eq.~\ref{equ:staterepresentation}, we can reconstruct general quantum states from the experimentally measured data. However, since the number of different measurement operators grows exponentially with the number of qubits, we need to prepare and measure the quantum states an exponential number of times to obtain the approximated distribution with small errors. For large quantum systems, this requirement is challenging to meet in current quantum devices. 
Here, we use a generative model to represent the exponentially complicated distributions. 

In the state reconstruction stage, we first train our generative model using the measurement results obtained from the experiments.
Once the model is trained,  we use it to reconstruct the quantum states. 
RNNs are utilized to learn the underlying distributions for high-dimensional sequential data.
Here, we treat the measurement result on a multiple-qubit system as a sequence of signals (a sequence of measurement outcomes for each qubit), and feed those signals into the RNNs sequentially.
The RNN-based generative model is trained to parameterize the distribution for those measurement outcomes. 
After the training procedure, the generative model is a classical representation of the quantum state, and we can draw random samples from the generative model to compute the desired physical properties of the quantum states. 
Due to the limited number of qubits in our experiment, we can also directly calculate the measurement distributions from the well-trained generative model and use the distributions to compute physical properties.

In this manuscript, we focus on three physical properties: 
\begin{enumerate}
  \item Classical fidelity: $F_C=\sum\left(P_{\text {model }} / P_{\text {exp }}\right)^{1 / 2}$;
  \item Quantum fidelity: $F_Q=\operatorname{Tr}\left(\sqrt{\rho_{\text {model }}^{1 / 2} \rho_{\text {exp }} \rho_{\text {model }}^{1 / 2}}\right)$;
  \item Two-body correlation: $C_{jk}=\text{Tr}\left(\rho_{\text{model}} {O}_{jk}\right)$,
  where $ {O}_{jk}= \sigma_x^{(j)}\otimes\sigma_x^{(k)},\  \sigma_z^{(j)}\otimes\sigma_z^{(k)}$ with $j,k=1,2,...,N$ and $N$ being the number of qubits.
\end{enumerate}

\begin{figure}
  \centering
  \includegraphics[width=0.8\textwidth]{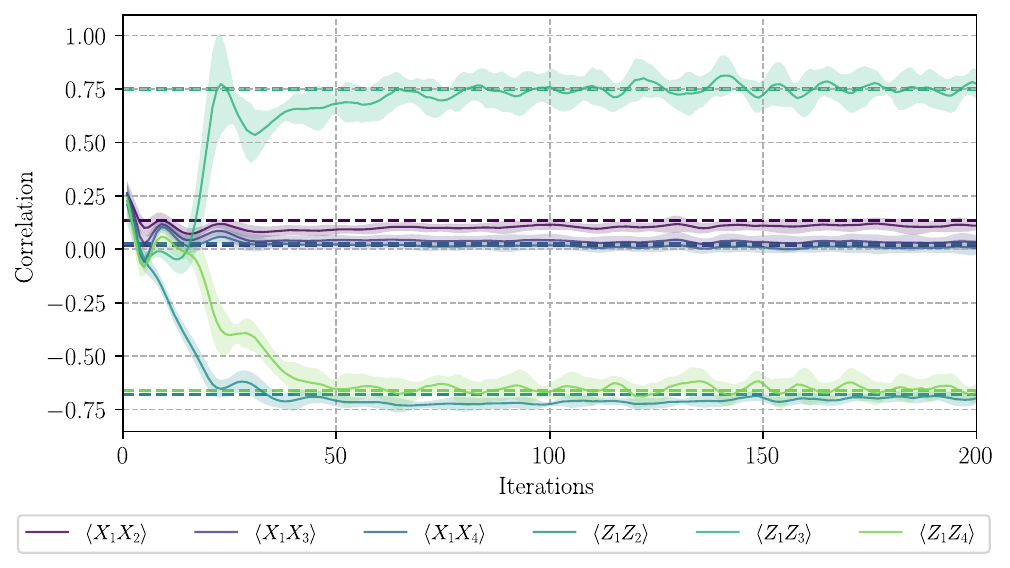}
  \caption{Reconstructed two-body $XX$ and $ZZ$ correlations without MLE correction. The corresponding experimental correlations computed using the QST method are shown as dashed lines.}
  \label{figs:ExpCorrsNoMLE}
\end{figure}

It is worth mentioning that, due to the inevitable error between the distributions of the experimentally prepared states and the learned states of the generative models, the quantum states reconstructed from the generative model can be illegitimate, i.e., the reconstructed density matrix has negative eigenvalues. In our work, we use the maximum-likelihood estimation (MLE) to correct the reconstructed density matrix. However, we note that this exponentially expensive correction is unnecessary if we are only concerned with the classical fidelity and the correlations (see Fig.~\ref{figs:ExpCorrsNoMLE}).

\subsection{Informationally complete POVM}
A POVM is called informationally complete if it consists of operators that span the space of all linear operators acting on the corresponding Hilbert space. Therefore, an informationally complete POVM on an $n$-qubit system must comprise at least $2^{2n}$ operators.
Since the density matrix of a quantum state is a Hermitian, semi-positive operator acting on the Hilbert space, the density matrices $\rho$ of both pure and mixed quantum states can be represented as linear combinations of operators from an informationally complete POVM.

In the main text, we primarily focus on the Pauli-4 POVM as defined in Eq.~\ref{eq:Pauli4}. 
This POVM is experimentally friendly since we can manually group three different outcomes of $\ket{1}\bra{1},\ \ket{-}\bra{-}$, and $\ket{r}\bra{r}$. The corresponding overlap matrix is 
$$
T_{\text {Pauli-4 }}=\frac{1}{9}\left(\begin{array}{cccc}
1 & 1 / 2 & 1 / 2 & 1 \\
1 / 2 & 1 & 1 / 2 & 1 \\
1 / 2 & 1 / 2 & 1 & 1 \\
1 & 1 & 1 & 6
\end{array}\right).
$$

We also use the Pauli-6 POVM to reconstruct quantum states, defined as $M_{\text {Pauli-6}}=\left\{M^{(0)}=\frac{1}{3}|0\rangle\langle 0|,\ M^{(1)}=\frac{1}{3}|+\rangle\langle+|, \ M^{(2)}=\frac{1}{3}| l\rangle\langle l|, \ M^{(3)}=\frac{1}{3}\{|1\rangle\langle 1|,\ M^{(4)}=|-\rangle\langle-|,\ \ M^{(5)}=|{r}\rangle\langle{r}|\right\}$). The corresponding overlap matrix is 
$$
T_{\text {Pauli-6 }}=\frac{1}{9}\left(\begin{array}{llllll}
1 & 0 & 1 / 2 & 1 / 2 & 1 / 2 & 1 / 2 \\
0 & 1 & 1 / 2 & 1 / 2 & 1 / 2 & 1 / 2 \\
1 / 2 & 1 / 2 & 1 & 0 & 1 / 2 & 1 / 2 \\
1 / 2 & 1 / 2 & 0 & 1 & 1 / 2 & 1 / 2 \\
1 / 2 & 1 / 2 & 1 / 2 & 1 / 2 & 1 & 0 \\
1 / 2 & 1 / 2 & 1 / 2 & 1 / 2 & 0 & 1
\end{array}\right).
$$
However, this matrix is not invertible, making it challenging to reconstruct quantum states using this POVM. To address this, we first learn the measurement distribution of the Pauli-6 POVM. Then, we combine the last three different outcomes to obtain a distribution of the Pauli-4 POVM. Finally, we reconstruct the measured quantum states from the Pauli-4 POVM distribution.

\subsection{Recurrent neural network}

\begin{algorithm}[t]
  \caption{Training RNN\label{alg:training}}
    \SetAlgoLined
    \KwIn{Learning rate $\epsilon$, training set $S$, threshold $\alpha$}
    \KwOut{Optimal parameters of RNN}
    $L\leftarrow +\infty$\;
    Randomly initialized $\bm{\theta}$\;
    \While{ $L>\alpha$}{
      $L\leftarrow 0$\;
      Randomly chosen state $h_0$\;
      \For{$S_k\in S$}{
        $\hat{S}_k\leftarrow\bm{0}$\;
        \For{$s_m\in S_k$}{
          $\hat{S}_k[m], h_m \leftarrow \text{RNN}(h_{m-1}, s_m; \bm{\theta})$\;
        }
        $L\leftarrow L+\text{Loss}(\hat{S}_k,S_k)$\;
      }
      $\bm{\theta}\leftarrow\bm{\theta} - \epsilon \nabla_{\bm{\theta}} L$\;
    }
    \KwRet{$\{P_k\}$}\;
\end{algorithm}

\begin{algorithm}[t]
  \caption{Computing learned distribution\label{alg:dis}}
    \SetAlgoLined
    \KwIn{Trained RNN, number of qubits $N$}
    \KwOut{Distribution \{$P_k$\}}
    Randomly chosen state $h_0$\;
    \For{$S_k\in$ all possible measurement results}{
      $P_k\leftarrow 1$\;
      \For{$s_m\in S_k$}{
        $p_m\leftarrow \text{RNN}(h_{m-1},s_m)$\;
        $P_k \leftarrow p_m*P_k$\;
      }
    }
    \KwRet{$\bm{\theta}$}\;
\end{algorithm}

RNNs are a family of artificial neural networks designed for processing sequential tasks. They leverage the idea of parameter sharing and can be extended and applied to sequential tasks, such as the stock price prediction, machine translation, etc. \cite{Salehinejad2018Recent,Sutskever2011Generating}. 
An RNN processes all inputs using the same neural network architecture and parameters sequentially. After processing each input, a hidden state with fixed length is passed forward, as shown in Fig.~1(c) of the main text. The hidden state contains information from signals processed before, thus influencing the outputs of the following signals. 
Mathematically, the output of an RNN can be represented as $y_k=f_1(h_{k-1},x_k)$ and the hidden state can be represented as $h_k=f_2(h_{k-1},x_k)$, where $x_{k}$ is the $k$-th input signal and $h_{k-1}$ is the $(k-1)$-th hidden state.

In this work, we use the RNNs as the network architecture for the generative model to reconstruct quantum states from the trained neural networks. 
In the training phase, the RNN takes the measurement results of the experimentally prepared quantum states as the input sequences, and minimizes the corresponding loss function (the negative cross-entropy function) to fit the underlying distributions $P_\text{exp}$ (see Algorithm~\ref{alg:training}). 
In the generating phase, the generative model takes a random outcome of the POVM as the first input, and outputs the marginal distribution of the measurement outcomes for the first qubit. We then draw a random sample from the distribution as the measurement result of the first qubit, and use it as the input to obtain the marginal distribution for the second qubit. Subsequently, we draw a random sample from the distribution as the measurement result of the second qubit. The above procedure is repeated until a complete measurement sequence of the experimental system is obtained. 
We can generate an ensemble of the measurement sequences and compute desired physical quantities. 
In our work, since the number of qubits is up to five, we can explicitly compute the distribution of all possible measurement sequences represented by the generative model (see Algorithm~\ref{alg:dis}), and then reconstruct the density matrices from those distributions using Eq.~\ref{equ:staterepresentation}. 
We note that this method is only practical for small systems. Due to the exponential growth of the Hilbert space, directly computing the distributions for large systems becomes impractical. 
In such cases, we can draw random samples from the generative model to compute the desired physical quantities.

\subsection{Maximum likelihood estimation}

In our experiments, the probabilities computed from the generative model can be unphysical due to inevitable small deviations of the experimentally prepared quantum states. Therefore, it is necessary to use a procedure that accounts for these errors.

We use the MLE method to correct the reconstructed density matrix from the distributions of the well-trained RNNs. Legitimate density matrices are semi-positive, Hermitian, and have a trace equal to one. We define the error between the measurement distributions as:
\begin{equation}
  L=\|P_\text{model}-P_\text{MLE}\|^2   .
\end{equation}
This error serves as the objective function, which is 
minimized with legitimate distribution $P_\text{MLE}$ by the MLE method. We implement the above procedure using an open-source package~\cite{Diamond2016CVXPY}.

\section{EXPERIMENTAL SETTING}

\subsection{Device parameters}

\begin{figure}[ht]
\includegraphics[scale=0.8]{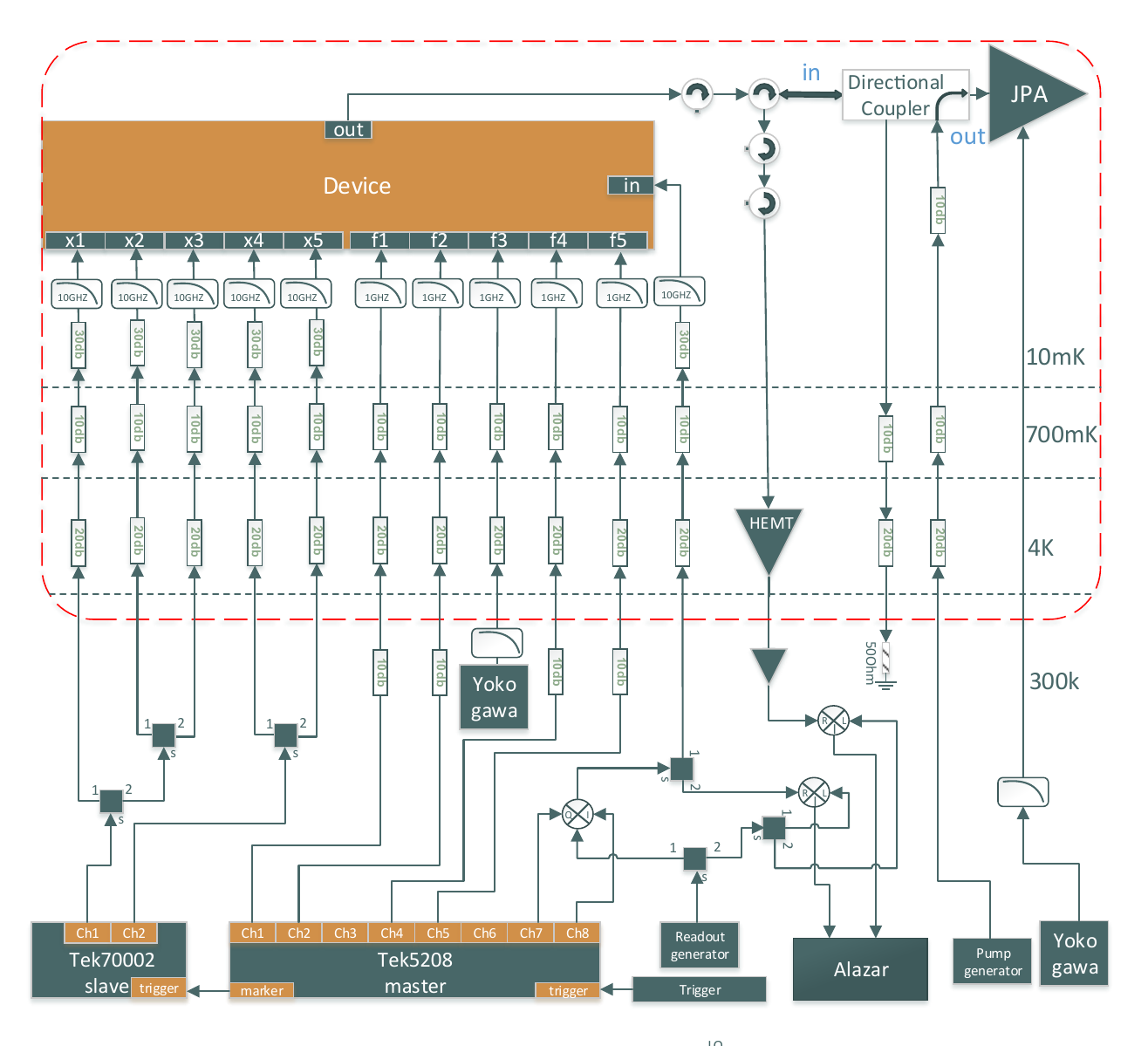}
\caption{Experimental setup.}
\label{figs:Experimental setup}
\end{figure}

The fabrication and calibration of our five-qubit device are the same as those in Ref.~\cite{Li2018Perfect}. The device parameters are shown in Table~\ref{Table:parameters}. The experimental setup is slightly different and shown in Fig.~\ref{figs:Experimental setup}. Each qubit has an independent flux control line and the control pulse is generated by arbitrary waveform generator (AWG) Tek5208, except for $Q_3$ whose flux is biased with a voltage source (Yokogawa). In our experiment, the frequency of $Q_3$ needs to be tuned lower and this will generally lead to a lower coherence time of $T_2$. Since Yokogawa has less voltage noise than Tek5208, we can acquire a higher $T_2$ of $Q_3$. The XY control pulse of each qubit is generated by a Tek70002 which has a large bandwidth about 8 GHz, enough to directly drive the qubits whose frequencies are arranged from 4.45~GHz to 5.16~GHz. 
The XY control lines of $Q_1$, $Q_2$, and $Q_3$ share one channel of Tek70002, and the XY control lines of $Q_4$ and $Q_5$ share the other channel of Tek70002. This would lead to XY crosstalk between the shared qubits. Through carefully separating each qubit frequency, the XY crosstalk can be suppressed to a low level.

\begin{figure}
\includegraphics{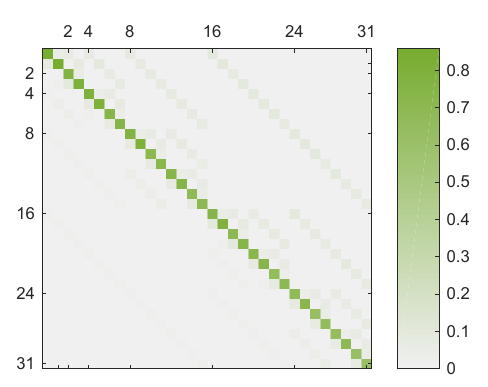}
\caption{Five qubit simultaneous readout matrix. The decimal numbers in the X and Y axes correspond to the binary number states. For example, the number 4 represents the state $\ket{00100}$.}
\label{figs:Readout_matrix_s2}
\end{figure}

The five qubits are simultaneously read out by using a common carrier from one generator which is modulated with five sideband frequencies generated with Tek5208. The readout signals are amplified by a Josephson parametric amplifier (JPA) with a bandwidth of about 260~MHz. This is enough to read out all five qubits simultaneously. The readout fidelities of the five qubits are shown in Table~\ref{Table:ReadoutFidelity}. The readout crosstalk is small and the readout matrix of the five qubits are measured as shown in Fig.~\ref{figs:Readout_matrix_s2}. 

The operating points of the five qubits have been optimized for a high-fidelity five-qubit GHZ state. Although all qubits at the sweet spot have higher $T_2$, the ZZ coupling between neighbouring qubits would lead to a large error on the GHZ state preparation and measurement. After enlarging the detuning frequency between neighboring qubits, the ZZ coupling strength $|\xi_j|$ becomes acceptable to prepare a high-fidelity GHZ state, as shown in Table~\ref{Table:parameters}.  

\begin{table*}
\caption{ {Device parameters.} }
\begin{tabular}{cp{0.9cm}<{\centering}p{0.9cm}<{\centering}p{0.9cm}<{\centering}p{0.9cm}<{\centering}p{0.9cm}<{\centering}p{0.9cm}<{\centering}p{0.9cm}<{\centering}p{0.9cm}<{\centering}p{0.9cm}<{\centering}p{0.9cm}<{\centering}}
&&&&&&&& \tabularnewline
\hline
\hline
Parameters &\multicolumn{2}{c}{$Q_1$}  &\multicolumn{2}{c}{$Q_2$} &\multicolumn{2}{c}{$Q_3$} &\multicolumn{2}{c}{$Q_4$} &\multicolumn{2}{c}{$Q_5$}\tabularnewline
\hline
Readout frequency (GHz) &\multicolumn{2}{c}{$6.8387$} &\multicolumn{2}{c}{$6.8634$} &\multicolumn{2}{c}{$6.879$} &\multicolumn{2}{c}{$6.9013$} &\multicolumn{2}{c}{$6.9191$} \tabularnewline
Qubit frequency (GHz) (sweet spot) &\multicolumn{2}{c}{$4.781$} &\multicolumn{2}{c}{$5.124$} &\multicolumn{2}{c}{$4.896$} &\multicolumn{2}{c}{$5.160$} &\multicolumn{2}{c}{$4.57$} \tabularnewline
$T_1$ ($\mu$s) (sweet spot) &\multicolumn{2}{c}{$21.54$} &\multicolumn{2}{c}{$25.18$} &\multicolumn{2}{c}{$21.00$} &\multicolumn{2}{c}{$27$} &\multicolumn{2}{c}{$24$} \tabularnewline
$T^*_2$ ($\mu$s) (sweet spot) &\multicolumn{2}{c}{$28.02$} &\multicolumn{2}{c}{$13.40$} &\multicolumn{2}{c}{$8.22$} &\multicolumn{2}{c}{$28.4$} &\multicolumn{2}{c}{$20.57$} \tabularnewline
Qubit frequency (GHz) (operating point) &\multicolumn{2}{c}{$4.615$} &\multicolumn{2}{c}{$5.0831$} &\multicolumn{2}{c}{$4.451$} &\multicolumn{2}{c}{$5.160$} &\multicolumn{2}{c}{$4.550$} \tabularnewline
$T_1$ ($\mu$s) (operating point) &\multicolumn{2}{c}{$20.174$} &\multicolumn{2}{c}{$19.27$} &\multicolumn{2}{c}{$15.47$} &\multicolumn{2}{c}{$22.9$}  &\multicolumn{2}{c}{$23.39$} \tabularnewline
$T^*_2$ ($\mu$s) (operating point) &\multicolumn{2}{c}{$2.37$} &\multicolumn{2}{c}{$6.99$} &\multicolumn{2}{c}{$3.45$} &\multicolumn{2}{{c}}{$28.4$} &\multicolumn{2}{{c}}{$9.18$} \tabularnewline
Average single-qubit gate fidelity &\multicolumn{2}{c}{$0.9980$} &\multicolumn{2}{c}{$0.9992$} &\multicolumn{2}{c}{$0.9987$} &\multicolumn{2}{{c}}{$0.9983$} &\multicolumn{2}{{c}}{$0.9991$} \tabularnewline
\hline
\hline
Neighboring qubits & &\multicolumn{2}{c}{$Q_1$-$Q_2$} &\multicolumn{2}{c}{$Q_2$-$Q_3$}& \multicolumn{2}{c}{$Q_3$-$Q_4$}&\multicolumn{2}{c}{$Q_4$-$Q_5$}& \tabularnewline

Capacitive coupling strength ${\rm g}_{j}/2\pi$ (MHz) & &\multicolumn{2}{c}{$16.68$} &\multicolumn{2}{c}{$17.50$} &\multicolumn{2}{c}{$17.52$} & \multicolumn{2}{c}{$17.62$} & \tabularnewline
ZZ coupling strength ${\rm |\xi}_{j}|/2\pi$ (MHz) (operating point) & &\multicolumn{2}{c}{$1.20$} &\multicolumn{2}{c}{$0.65$} &\multicolumn{2}{c}{$0.59$} & \multicolumn{2}{c}{$0.58$} & \tabularnewline
Controlled-phase gate fidelity  & &\multicolumn{2}{c}{$0.967$} &\multicolumn{2}{c}{$0.957$} &\multicolumn{2}{c}{$0.970$} & \multicolumn{2}{c}{$0.99$} & \tabularnewline
\hline
\hline
\end{tabular} \vspace{-6pt}
\label{Table:parameters}
\end{table*}

\begin{table}
\caption{The readout fidelities for each qubit. $F_g$ is the measured fidelity for an inital thermal steady state and $F_e$ is for an initial thermal steady state followed by a $\pi$ rotation gate.} 
\begin{center}
\begin{tabular}{p{1.5cm}<{\centering}p{1.5cm}<{\centering}p{1.5cm}<{\centering}p{1.5cm}<{\centering}p{1.5cm}<{\centering}p{1.5cm}<{\centering}}
\hline
\hline
\centering
       & $Q_1$  & $Q_2   $ & $Q_3$ & $Q_4$ & $Q_5$\\
\hline
$F_{{\rm g}}$  & 0.990  & 0.985 & 0.973 & 0.988 & 0.985 \\
$F_{{\rm e}}$   & 0.899  & 0.933 & 0.922 & 0.917 & 0.918   \\
\hline
\hline
\end{tabular} \vspace{8pt}\\
\label{Table:ReadoutFidelity}
\end{center}
\end{table}

\begin{figure}[ht]
\includegraphics[scale=1]{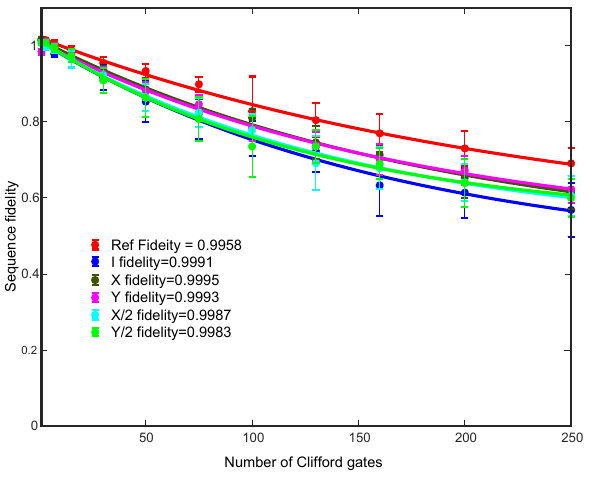}
\caption{Randomized benchmarking of single-qubit gates on $Q_3$.}
\label{figs:SingleQubitRB}
\end{figure}

\begin{figure}[ht]
\includegraphics[scale=1]{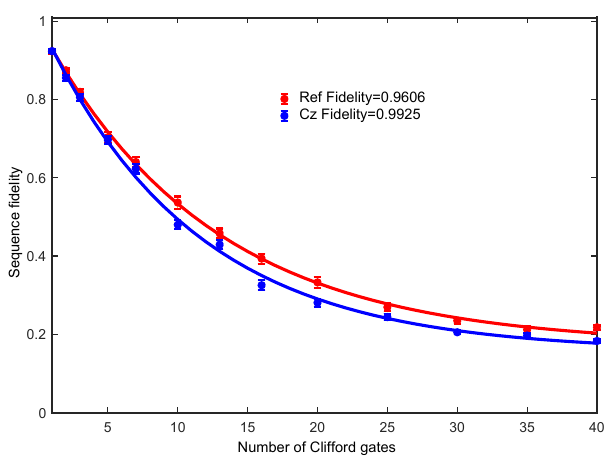}
\caption{Randomized benchmarking of the controlled-phase gate on $Q_4$ and $Q_5$.}
\label{figs:Cz_RB}
\end{figure}

\subsection{Single-qubit gate and controlled-phase gate}
The fidelity of single-qubit gates is limited by many factors, such as Z control crosstalk, microwave XY control crosstalk, and ZZ coupling between neighboring qubits. To address these limitations, we first tune the frequencies of neighboring qubits by biasing certain qubits away from their sweet spots to reduce the microwave XY control crosstalk and ZZ couplings, sacrificing the phase coherence times. Additionally, we apply the deconvolution technique to suppress the ripple after the controlled-phase flux pulse. Furthermore, we calibrate the delay between the Z control pulse and XY control pulse to further reduce any residual ripple. After careful calibration of the unwanted ripples, we characterize the fidelity of each single-qubit gate at the chosen frequency using the Clifford-based randomized benchmarking, with $Q_3$ serving as an example shown in Fig.~\ref{figs:SingleQubitRB}. On average, the single-qubit gate fidelity of all qubits is about $99.87\%$. Additionally, we assess the ZZ coupling and implement simultaneous randomized benchmarking between nearest neighboring qubits. We find that the average simultaneous single-qubit gate fidelity is lowered by about $0.3\%$. Detailed benchmarking results, operation frequency of each qubit, and ZZ coupling strength between neighboring qubits can be found in Table~\ref{Table:parameters}. 

We implement the two-qubit controlled-phase gate by tuning the frequency of one of the neighboring qubits to bring the two qubits into $\ket{11}$ and $\ket{02}$ avoided-crossing point. There are two methods to implement this goal based on frequency-tunable qubits. One is to quickly bring the system into resonance via a rectangular flux pulse and wait for a certain time, followed by quickly moving the system back to the idle point. The states of two qubits will remain the same under the quickly changed Hamiltonian, but a total geometrical $\pi$ phase can be accumulated during the interaction time. However, the finite sample rate of the AWG makes this quick flux pulse imperfect and would lead to a population leakage out of the computational subspace. The other method is to tune the qubit along a fast adiabatic trajectory, which does not require a change faster than the sample rate of the AWG. To get a high-fidelity controlled-phase gate, this trajectory has been optimized by system Hamiltonian using Nelder-Mead algorithm~\cite{martinis2014fast}. This method can suppress the leakage error and meanwhile keep a fast gate time. We adopt the latter method and use the two-qubit Clifford-based randomized benchmarking to characterize the performance of the controlled-phase gate. The results can be found in Table~\ref{Table:parameters}. The controlled-phase gate fidelity between $Q_4$ and $Q_5$ reaches up to $99.2\%$, as shown in Fig.~\ref{figs:Cz_RB}. The gate fidelities of other neighboring qubit pairs are lower due to the low coherence of $Q_1$ and $Q_3$, as well as the unwanted two-level systems near the trajectory of the $Q_2$ flux pulse.

\subsection{Pauli-4 POVM for a random state}
\label{apps:RandomState}

\begin{figure*}[t]
\includegraphics[width=\textwidth]{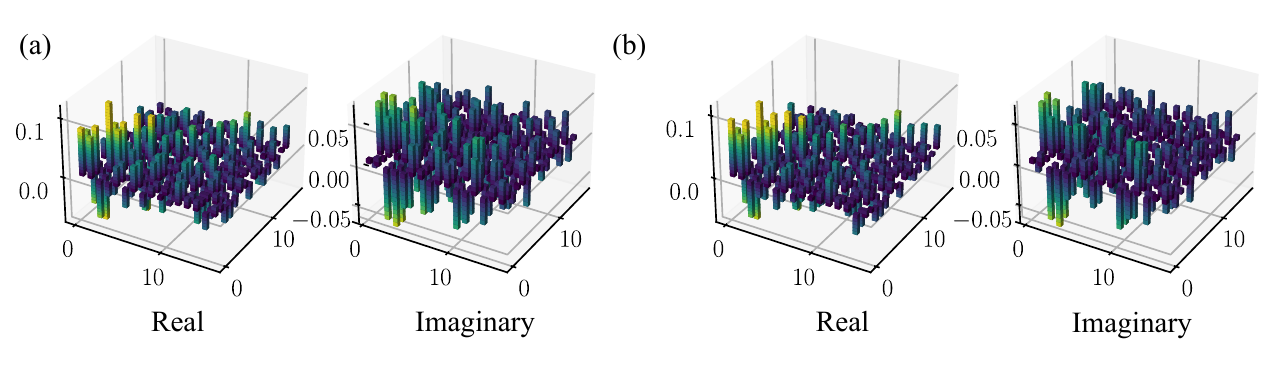}
\caption{ {Reconstructed random states.} 
(a) The density matrix of the randomly chosen state reconstructed directly from experimental QST. 
(b) The density matrix reconstructed by maximizing the classical fidelity ($F_C>99\%$). }
\label{figs:ExpStateArbitary}
\end{figure*}

\begin{figure}[t]
\includegraphics[width=\textwidth]{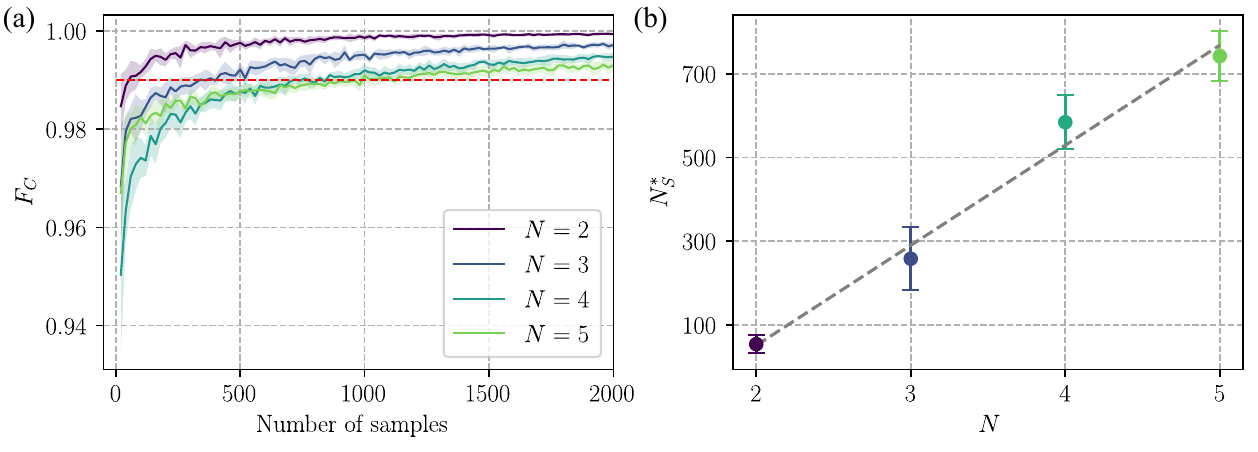}
\caption{ {Reconstructing random states using Pauli-4 POVM.} 
(a) The average classical fidelity as a function of the necessary number of samples $N_s$ performed on the random state with $N = 2,\,3,\,4,\,5$ qubits. The red dashed line denotes the threshold of $F_C=99\%$ for benchmarking the necessary number of samples $N_s$. We repeat this process $10$ times for each qubit number $N$ by randomly choosing the samples from the experimental data. The transparent filled regions indicate the error bars. 
(b) The necessary number of samples $N_s$ for reconstructing random states with $F_C=99\%$ scales linearly with qubit number $N$. The gray dashed line denotes a linear fit of the experimental data.}
\label{figs:ExpScalingArbitary}
\end{figure}

In this session, we focus on a random state with no special physical structure and reconstruct such a randomly chosen quantum state using the generative model.
We randomly prepare a quantum state by tuning all the qubit frequencies nearly equal and allowing the system to evolve for a certain time. After the preparation, we measure the state using aforementioned informationally complete POVM and carry out the tomography process.
For such a state, we do not predict the density matrix in advance. This final density matrix for the random state is constructed by quantum state tomography (QST), as shown in Fig.~\ref{figs:ExpStateArbitary}(a). 
We perform the generative-model-based tomography to reconstruct the random state with $N=4$ qubits, obtaining the reconstructed density matrix as shown in Fig.~\ref{figs:ExpStateArbitary} (b). 
We find the achieved classical fidelity $F_C$ is larger than $99\%$. Furthermore, we carry out this tomography process for similarly generated random states with different numbers of qubits [see Fig.~\ref{figs:ExpScalingArbitary}(a)], and observe that the necessary number of measurements to achieve $F_C>99\%$ scales approximately linearly, as illustrated in Fig.~\ref{figs:ExpScalingArbitary}(b).

\subsection{Pauli-6 POVM for GHZ states}

\begin{figure}[t]
\includegraphics[width=\textwidth]{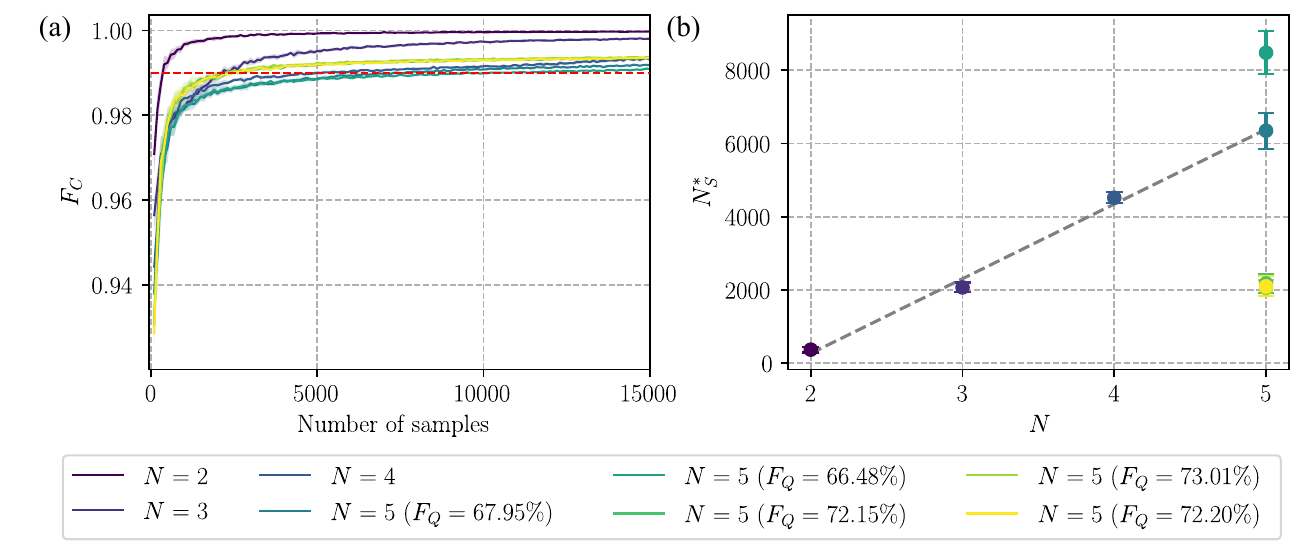}
\caption{Reconstructing GHZ states using Pauli-6 POVM. We remark that the quantum state with $F_Q=72.20\%$ is that used in main text ($F_Q=89.4\%$ in FIG.~2, estimated after Bayes correction and maximum likelihood estimation).
(a) The average classical fidelity as a function of the necessary number of samples for GHZ states with $N = 2,\, 3,\, 4,\, 5$ qubits. The red dashed line denotes the threshold of $F_C=99\%$. We repeat this process $10$ times by randomly choosing the samples from the experimental data. 
(b) The necessary number of samples $N_s$ for reconstructing GHZ state with $F_C=99\%$. The gray dashed line denotes a linear fit for $N=2,\,3,\,4$, and 5-qubit GHZ states with fidelities of $F_C=99\%$. We also show the necessary number of samples for five-qubit GHZ states prepared with different experimental settings.}
\label{figs:ExpScalingGHZ6Pauli}
\end{figure}

We also use the Pauli-6 POVM to carry out the generative-model-based tomography for reconstructing the GHZ states. Pauli-6 POVM has more measurement operators and the corresponding distribution is redundant to represent a quantum state.
Thus, to avoid possible misleading, we use the classical fidelity between the learned probability and the measured probability (without Bayes correction and maximum likelihood estimation) as an indicator and show the averaged learning curves in Fig.~\ref{figs:ExpScalingGHZ6Pauli}(a). We observe that the required number of samples to reach a threshold of $F_C>99\%$ is much larger than that for Pauli-4 POVM [see Fig.~\ref{figs:ExpScalingGHZ6Pauli}(b)]. 

\begin{figure}[t]
  \includegraphics[width=\textwidth]{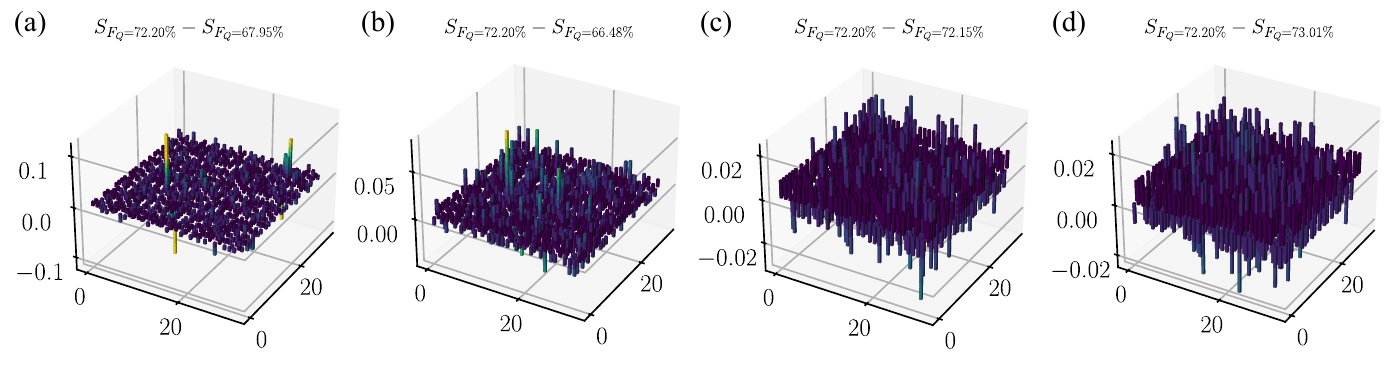}
  \caption{ {State difference of five-qubit GHZ states.} Here, we fix our reference state as the experiment state used in main text, whose quantum fidelity is $F_Q=72.20\%$, and show its difference with density matrices of other experimentally prepared states.}
  \label{figs:StateDiff}
  \end{figure}

Furthermore, we find that the necessary number of samples is affected by the experimental states. We prepare several five-qubit GHZ states with different experimental settings and different fidelities. The learning results show that the necessary numbers of samples to reach $F_C>99\%$ are different, as shown in Fig.~\ref{figs:ExpScalingGHZ6Pauli} (b). For the five-qubit GHZ state prepared using the experimental setting similar with that for $N=2,\,3,\,4$ (fidelity is about $67.95\%$), we find the scaling is approximately linear. 
We conjecture that the distinction comes from the subtle differences between quantum states encountered in practical quantum devices (see Fig.~\ref{figs:StateDiff}) and the redundancy of the Pauli-6 POVM. However, this problem requires further investigation.

\begin{thebibliography}{44}%
\makeatletter
\providecommand \@ifxundefined [1]{%
 \@ifx{#1\undefined}
}%
\providecommand \@ifnum [1]{%
 \ifnum #1\expandafter \@firstoftwo
 \else \expandafter \@secondoftwo
 \fi
}%
\providecommand \@ifx [1]{%
 \ifx #1\expandafter \@firstoftwo
 \else \expandafter \@secondoftwo
 \fi
}%
\providecommand \natexlab [1]{#1}%
\providecommand \enquote  [1]{``#1''}%
\providecommand \bibnamefont  [1]{#1}%
\providecommand \bibfnamefont [1]{#1}%
\providecommand \citenamefont [1]{#1}%
\providecommand \href@noop [0]{\@secondoftwo}%
\providecommand \href [0]{\begingroup \@sanitize@url \@href}%
\providecommand \@href[1]{\@@startlink{#1}\@@href}%
\providecommand \@@href[1]{\endgroup#1\@@endlink}%
\providecommand \@sanitize@url [0]{\catcode `\\12\catcode `\$12\catcode
  `\&12\catcode `\#12\catcode `\^12\catcode `\_12\catcode `\%12\relax}%
\providecommand \@@startlink[1]{}%
\providecommand \@@endlink[0]{}%
\providecommand \url  [0]{\begingroup\@sanitize@url \@url }%
\providecommand \@url [1]{\endgroup\@href {#1}{\urlprefix }}%
\providecommand \urlprefix  [0]{URL }%
\providecommand \Eprint [0]{\href }%
\providecommand \doibase [0]{http://dx.doi.org/}%
\providecommand \selectlanguage [0]{\@gobble}%
\providecommand \bibinfo  [0]{\@secondoftwo}%
\providecommand \bibfield  [0]{\@secondoftwo}%
\providecommand \translation [1]{[#1]}%
\providecommand \BibitemOpen [0]{}%
\providecommand \bibitemStop [0]{}%
\providecommand \bibitemNoStop [0]{.\EOS\space}%
\providecommand \EOS [0]{\spacefactor3000\relax}%
\providecommand \BibitemShut  [1]{\csname bibitem#1\endcsname}%
\let\auto@bib@innerbib\@empty
\bibitem [{\citenamefont {Harrigan}\ \emph {et~al.}(2021)\citenamefont
  {Harrigan}, \citenamefont {Sung}, \citenamefont {Neeley}, \citenamefont
  {Satzinger}, \citenamefont {Arute}, \citenamefont {Arya}, \citenamefont
  {Atalaya}, \citenamefont {Bardin}, \citenamefont {Barends}, \citenamefont
  {Boixo} \emph {et~al.}}]{Harrigan2021Quantum}%
  \BibitemOpen
  \bibfield  {author} {\bibinfo {author} {\bibfnamefont {M.~P.}\ \bibnamefont
  {Harrigan}}, \bibinfo {author} {\bibfnamefont {K.~J.}\ \bibnamefont {Sung}},
  \bibinfo {author} {\bibfnamefont {M.}~\bibnamefont {Neeley}}, \bibinfo
  {author} {\bibfnamefont {K.~J.}\ \bibnamefont {Satzinger}}, \bibinfo {author}
  {\bibfnamefont {F.}~\bibnamefont {Arute}}, \bibinfo {author} {\bibfnamefont
  {K.}~\bibnamefont {Arya}}, \bibinfo {author} {\bibfnamefont {J.}~\bibnamefont
  {Atalaya}}, \bibinfo {author} {\bibfnamefont {J.~C.}\ \bibnamefont {Bardin}},
  \bibinfo {author} {\bibfnamefont {R.}~\bibnamefont {Barends}}, \bibinfo
  {author} {\bibfnamefont {S.}~\bibnamefont {Boixo}},  \emph {et~al.},\ }\href
  {\doibase 10.1038/s41567-020-01105-y} {\bibfield  {journal} {\bibinfo
  {journal} {Nat. Phys.}\ }\textbf {\bibinfo {volume} {17}},\ \bibinfo {pages}
  {332} (\bibinfo {year} {2021})}\BibitemShut {NoStop}%
\bibitem [{\citenamefont {Kandala}\ \emph {et~al.}(2017)\citenamefont
  {Kandala}, \citenamefont {Mezzacapo}, \citenamefont {Temme}, \citenamefont
  {Takita}, \citenamefont {Brink}, \citenamefont {Chow},\ and\ \citenamefont
  {Gambetta}}]{Kandala2017Hardwareefficient}%
  \BibitemOpen
  \bibfield  {author} {\bibinfo {author} {\bibfnamefont {A.}~\bibnamefont
  {Kandala}}, \bibinfo {author} {\bibfnamefont {A.}~\bibnamefont {Mezzacapo}},
  \bibinfo {author} {\bibfnamefont {K.}~\bibnamefont {Temme}}, \bibinfo
  {author} {\bibfnamefont {M.}~\bibnamefont {Takita}}, \bibinfo {author}
  {\bibfnamefont {M.}~\bibnamefont {Brink}}, \bibinfo {author} {\bibfnamefont
  {J.~M.}\ \bibnamefont {Chow}}, \ and\ \bibinfo {author} {\bibfnamefont
  {J.~M.}\ \bibnamefont {Gambetta}},\ }\href {\doibase 10.1038/nature23879}
  {\bibfield  {journal} {\bibinfo  {journal} {Nature}\ }\textbf {\bibinfo
  {volume} {549}},\ \bibinfo {pages} {242} (\bibinfo {year}
  {2017})}\BibitemShut {NoStop}%
\bibitem [{\citenamefont {Arute}\ \emph {et~al.}(2019)\citenamefont {Arute},
  \citenamefont {Arya}, \citenamefont {Babbush}, \citenamefont {Bacon},
  \citenamefont {Bardin}, \citenamefont {Barends}, \citenamefont {Biswas},
  \citenamefont {Boixo}, \citenamefont {Brandao}, \citenamefont {Buell} \emph
  {et~al.}}]{Arute2019Quantum}%
  \BibitemOpen
  \bibfield  {author} {\bibinfo {author} {\bibfnamefont {F.}~\bibnamefont
  {Arute}}, \bibinfo {author} {\bibfnamefont {K.}~\bibnamefont {Arya}},
  \bibinfo {author} {\bibfnamefont {R.}~\bibnamefont {Babbush}}, \bibinfo
  {author} {\bibfnamefont {D.}~\bibnamefont {Bacon}}, \bibinfo {author}
  {\bibfnamefont {J.~C.}\ \bibnamefont {Bardin}}, \bibinfo {author}
  {\bibfnamefont {R.}~\bibnamefont {Barends}}, \bibinfo {author} {\bibfnamefont
  {R.}~\bibnamefont {Biswas}}, \bibinfo {author} {\bibfnamefont
  {S.}~\bibnamefont {Boixo}}, \bibinfo {author} {\bibfnamefont {F.~G. S.~L.}\
  \bibnamefont {Brandao}}, \bibinfo {author} {\bibfnamefont {D.~A.}\
  \bibnamefont {Buell}},  \emph {et~al.},\ }\href {\doibase
  10.1038/s41586-019-1666-5} {\bibfield  {journal} {\bibinfo  {journal}
  {Nature}\ }\textbf {\bibinfo {volume} {574}},\ \bibinfo {pages} {505}
  (\bibinfo {year} {2019})}\BibitemShut {NoStop}%
\bibitem [{\citenamefont {Song}\ \emph {et~al.}(2019)\citenamefont {Song},
  \citenamefont {Xu}, \citenamefont {Li}, \citenamefont {Zhang}, \citenamefont
  {Zhang}, \citenamefont {Liu}, \citenamefont {Guo}, \citenamefont {Wang},
  \citenamefont {Ren}, \citenamefont {Hao} \emph
  {et~al.}}]{Song2019Generation}%
  \BibitemOpen
  \bibfield  {author} {\bibinfo {author} {\bibfnamefont {C.}~\bibnamefont
  {Song}}, \bibinfo {author} {\bibfnamefont {K.}~\bibnamefont {Xu}}, \bibinfo
  {author} {\bibfnamefont {H.}~\bibnamefont {Li}}, \bibinfo {author}
  {\bibfnamefont {Y.-R.}\ \bibnamefont {Zhang}}, \bibinfo {author}
  {\bibfnamefont {X.}~\bibnamefont {Zhang}}, \bibinfo {author} {\bibfnamefont
  {W.}~\bibnamefont {Liu}}, \bibinfo {author} {\bibfnamefont {Q.}~\bibnamefont
  {Guo}}, \bibinfo {author} {\bibfnamefont {Z.}~\bibnamefont {Wang}}, \bibinfo
  {author} {\bibfnamefont {W.}~\bibnamefont {Ren}}, \bibinfo {author}
  {\bibfnamefont {J.}~\bibnamefont {Hao}},  \emph {et~al.},\ }\href {\doibase
  10.1126/science.aay0600} {\bibfield  {journal} {\bibinfo  {journal}
  {Science}\ }\textbf {\bibinfo {volume} {365}},\ \bibinfo {pages} {574}
  (\bibinfo {year} {2019})}\BibitemShut {NoStop}%
\bibitem [{\citenamefont {Yan}\ \emph {et~al.}(2019)\citenamefont {Yan},
  \citenamefont {Zhang}, \citenamefont {Gong}, \citenamefont {Wu},
  \citenamefont {Zheng}, \citenamefont {Li}, \citenamefont {Wang},
  \citenamefont {Liang}, \citenamefont {Lin}, \citenamefont {Xu} \emph
  {et~al.}}]{Yan2019Strongly}%
  \BibitemOpen
  \bibfield  {author} {\bibinfo {author} {\bibfnamefont {Z.}~\bibnamefont
  {Yan}}, \bibinfo {author} {\bibfnamefont {Y.-R.}\ \bibnamefont {Zhang}},
  \bibinfo {author} {\bibfnamefont {M.}~\bibnamefont {Gong}}, \bibinfo {author}
  {\bibfnamefont {Y.}~\bibnamefont {Wu}}, \bibinfo {author} {\bibfnamefont
  {Y.}~\bibnamefont {Zheng}}, \bibinfo {author} {\bibfnamefont
  {S.}~\bibnamefont {Li}}, \bibinfo {author} {\bibfnamefont {C.}~\bibnamefont
  {Wang}}, \bibinfo {author} {\bibfnamefont {F.}~\bibnamefont {Liang}},
  \bibinfo {author} {\bibfnamefont {J.}~\bibnamefont {Lin}}, \bibinfo {author}
  {\bibfnamefont {Y.}~\bibnamefont {Xu}},  \emph {et~al.},\ }\href {\doibase
  10.1126/science.aaw1611} {\bibfield  {journal} {\bibinfo  {journal}
  {Science}\ }\textbf {\bibinfo {volume} {364}},\ \bibinfo {pages} {753}
  (\bibinfo {year} {2019})}\BibitemShut {NoStop}%
\bibitem [{\citenamefont {Ma}\ \emph {et~al.}(2019)\citenamefont {Ma},
  \citenamefont {Saxberg}, \citenamefont {Owens}, \citenamefont {Leung},
  \citenamefont {Lu}, \citenamefont {Simon},\ and\ \citenamefont
  {Schuster}}]{Ma2019Dissipatively}%
  \BibitemOpen
  \bibfield  {author} {\bibinfo {author} {\bibfnamefont {R.}~\bibnamefont
  {Ma}}, \bibinfo {author} {\bibfnamefont {B.}~\bibnamefont {Saxberg}},
  \bibinfo {author} {\bibfnamefont {C.}~\bibnamefont {Owens}}, \bibinfo
  {author} {\bibfnamefont {N.}~\bibnamefont {Leung}}, \bibinfo {author}
  {\bibfnamefont {Y.}~\bibnamefont {Lu}}, \bibinfo {author} {\bibfnamefont
  {J.}~\bibnamefont {Simon}}, \ and\ \bibinfo {author} {\bibfnamefont {D.~I.}\
  \bibnamefont {Schuster}},\ }\href {\doibase 10.1038/s41586-019-0897-9}
  {\bibfield  {journal} {\bibinfo  {journal} {Nature}\ }\textbf {\bibinfo
  {volume} {566}},\ \bibinfo {pages} {51} (\bibinfo {year} {2019})}\BibitemShut
  {NoStop}%
\bibitem [{\citenamefont {Koll{\'a}r}\ \emph {et~al.}(2019)\citenamefont
  {Koll{\'a}r}, \citenamefont {Fitzpatrick},\ and\ \citenamefont
  {Houck}}]{Kollar2019Hyperbolic}%
  \BibitemOpen
  \bibfield  {author} {\bibinfo {author} {\bibfnamefont {A.~J.}\ \bibnamefont
  {Koll{\'a}r}}, \bibinfo {author} {\bibfnamefont {M.}~\bibnamefont
  {Fitzpatrick}}, \ and\ \bibinfo {author} {\bibfnamefont {A.~A.}\ \bibnamefont
  {Houck}},\ }\href {\doibase 10.1038/s41586-019-1348-3} {\bibfield  {journal}
  {\bibinfo  {journal} {Nature}\ }\textbf {\bibinfo {volume} {571}},\ \bibinfo
  {pages} {45} (\bibinfo {year} {2019})}\BibitemShut {NoStop}%
\bibitem [{\citenamefont {Andersen}\ \emph {et~al.}(2020)\citenamefont
  {Andersen}, \citenamefont {Remm}, \citenamefont {Lazar}, \citenamefont
  {Krinner}, \citenamefont {Lacroix}, \citenamefont {Norris}, \citenamefont
  {Gabureac}, \citenamefont {Eichler},\ and\ \citenamefont
  {Wallraff}}]{Andersen2020Repeated}%
  \BibitemOpen
  \bibfield  {author} {\bibinfo {author} {\bibfnamefont {C.~K.}\ \bibnamefont
  {Andersen}}, \bibinfo {author} {\bibfnamefont {A.}~\bibnamefont {Remm}},
  \bibinfo {author} {\bibfnamefont {S.}~\bibnamefont {Lazar}}, \bibinfo
  {author} {\bibfnamefont {S.}~\bibnamefont {Krinner}}, \bibinfo {author}
  {\bibfnamefont {N.}~\bibnamefont {Lacroix}}, \bibinfo {author} {\bibfnamefont
  {G.~J.}\ \bibnamefont {Norris}}, \bibinfo {author} {\bibfnamefont
  {M.}~\bibnamefont {Gabureac}}, \bibinfo {author} {\bibfnamefont
  {C.}~\bibnamefont {Eichler}}, \ and\ \bibinfo {author} {\bibfnamefont
  {A.}~\bibnamefont {Wallraff}},\ }\href {\doibase 10.1038/s41567-020-0920-y}
  {\bibfield  {journal} {\bibinfo  {journal} {Nat. Phys.}\ }\textbf {\bibinfo
  {volume} {16}},\ \bibinfo {pages} {875} (\bibinfo {year} {2020})}\BibitemShut
  {NoStop}%
\bibitem [{\citenamefont {Jurcevic}\ \emph {et~al.}(2021)\citenamefont
  {Jurcevic}, \citenamefont {{Javadi-Abhari}}, \citenamefont {Bishop},
  \citenamefont {Lauer}, \citenamefont {Bogorin}, \citenamefont {Brink},
  \citenamefont {Capelluto}, \citenamefont {G{\"u}nl{\"u}k}, \citenamefont
  {Itoko}, \citenamefont {Kanazawa} \emph
  {et~al.}}]{Jurcevic2021Demonstration}%
  \BibitemOpen
  \bibfield  {author} {\bibinfo {author} {\bibfnamefont {P.}~\bibnamefont
  {Jurcevic}}, \bibinfo {author} {\bibfnamefont {A.}~\bibnamefont
  {{Javadi-Abhari}}}, \bibinfo {author} {\bibfnamefont {L.~S.}\ \bibnamefont
  {Bishop}}, \bibinfo {author} {\bibfnamefont {I.}~\bibnamefont {Lauer}},
  \bibinfo {author} {\bibfnamefont {D.~F.}\ \bibnamefont {Bogorin}}, \bibinfo
  {author} {\bibfnamefont {M.}~\bibnamefont {Brink}}, \bibinfo {author}
  {\bibfnamefont {L.}~\bibnamefont {Capelluto}}, \bibinfo {author}
  {\bibfnamefont {O.}~\bibnamefont {G{\"u}nl{\"u}k}}, \bibinfo {author}
  {\bibfnamefont {T.}~\bibnamefont {Itoko}}, \bibinfo {author} {\bibfnamefont
  {N.}~\bibnamefont {Kanazawa}},  \emph {et~al.},\ }\href {\doibase
  10.1088/2058-9565/abe519} {\bibfield  {journal} {\bibinfo  {journal} {Quantum
  Sci. Technol.}\ }\textbf {\bibinfo {volume} {6}},\ \bibinfo {pages} {025020}
  (\bibinfo {year} {2021})}\BibitemShut {NoStop}%
\bibitem [{\citenamefont {{GOOGLE AI QUANTUM AND COLLABORATORS}}\ \emph
  {et~al.}(2020)\citenamefont {{GOOGLE AI QUANTUM AND COLLABORATORS}},
  \citenamefont {Arute}, \citenamefont {Arya}, \citenamefont {Babbush},
  \citenamefont {Bacon}, \citenamefont {Bardin}, \citenamefont {Barends},
  \citenamefont {Boixo}, \citenamefont {Broughton}, \citenamefont {Buckley}
  \emph {et~al.}}]{GOOGLEAIQUANTUMANDCOLLABORATORS2020HartreeFock}%
  \BibitemOpen
  \bibfield  {author} {\bibinfo {author} {\bibnamefont {{GOOGLE AI QUANTUM AND
  COLLABORATORS}}}, \bibinfo {author} {\bibfnamefont {F.}~\bibnamefont
  {Arute}}, \bibinfo {author} {\bibfnamefont {K.}~\bibnamefont {Arya}},
  \bibinfo {author} {\bibfnamefont {R.}~\bibnamefont {Babbush}}, \bibinfo
  {author} {\bibfnamefont {D.}~\bibnamefont {Bacon}}, \bibinfo {author}
  {\bibfnamefont {J.~C.}\ \bibnamefont {Bardin}}, \bibinfo {author}
  {\bibfnamefont {R.}~\bibnamefont {Barends}}, \bibinfo {author} {\bibfnamefont
  {S.}~\bibnamefont {Boixo}}, \bibinfo {author} {\bibfnamefont
  {M.}~\bibnamefont {Broughton}}, \bibinfo {author} {\bibfnamefont {B.~B.}\
  \bibnamefont {Buckley}},  \emph {et~al.},\ }\href {\doibase
  10.1126/science.abb9811} {\bibfield  {journal} {\bibinfo  {journal}
  {Science}\ }\textbf {\bibinfo {volume} {369}},\ \bibinfo {pages} {1084}
  (\bibinfo {year} {2020})}\BibitemShut {NoStop}%
\bibitem [{\citenamefont {Bluvstein}\ \emph {et~al.}(2024)\citenamefont
  {Bluvstein}, \citenamefont {Evered}, \citenamefont {Geim}, \citenamefont
  {Li}, \citenamefont {Zhou}, \citenamefont {Manovitz}, \citenamefont {Ebadi},
  \citenamefont {Cain}, \citenamefont {Kalinowski}, \citenamefont {Hangleiter}
  \emph {et~al.}}]{Bluvstein2024Logical}%
  \BibitemOpen
  \bibfield  {author} {\bibinfo {author} {\bibfnamefont {D.}~\bibnamefont
  {Bluvstein}}, \bibinfo {author} {\bibfnamefont {S.~J.}\ \bibnamefont
  {Evered}}, \bibinfo {author} {\bibfnamefont {A.~A.}\ \bibnamefont {Geim}},
  \bibinfo {author} {\bibfnamefont {S.~H.}\ \bibnamefont {Li}}, \bibinfo
  {author} {\bibfnamefont {H.}~\bibnamefont {Zhou}}, \bibinfo {author}
  {\bibfnamefont {T.}~\bibnamefont {Manovitz}}, \bibinfo {author}
  {\bibfnamefont {S.}~\bibnamefont {Ebadi}}, \bibinfo {author} {\bibfnamefont
  {M.}~\bibnamefont {Cain}}, \bibinfo {author} {\bibfnamefont {M.}~\bibnamefont
  {Kalinowski}}, \bibinfo {author} {\bibfnamefont {D.}~\bibnamefont
  {Hangleiter}},  \emph {et~al.},\ }\href {\doibase 10.1038/s41586-023-06927-3}
  {\bibfield  {journal} {\bibinfo  {journal} {Nature}\ }\textbf {\bibinfo
  {volume} {626}},\ \bibinfo {pages} {58} (\bibinfo {year} {2024})}\BibitemShut
  {NoStop}%
\bibitem [{\citenamefont {Hawkes}(2003)}]{Hawkes2003Advances}%
  \BibitemOpen
  \bibfield  {author} {\bibinfo {author} {\bibfnamefont {P.~W.}\ \bibnamefont
  {Hawkes}},\ }\href@noop {} {\emph {\bibinfo {title} {Advances in {{Imaging}}
  and {{Electron Physics}}}}}\ (\bibinfo  {publisher} {Elsevier},\ \bibinfo
  {year} {2003})\BibitemShut {NoStop}%
\bibitem [{\citenamefont {Cramer}\ \emph {et~al.}(2010)\citenamefont {Cramer},
  \citenamefont {Plenio}, \citenamefont {Flammia}, \citenamefont {Somma},
  \citenamefont {Gross}, \citenamefont {Bartlett}, \citenamefont
  {{Landon-Cardinal}}, \citenamefont {Poulin},\ and\ \citenamefont
  {Liu}}]{Cramer2010Efficient}%
  \BibitemOpen
  \bibfield  {author} {\bibinfo {author} {\bibfnamefont {M.}~\bibnamefont
  {Cramer}}, \bibinfo {author} {\bibfnamefont {M.~B.}\ \bibnamefont {Plenio}},
  \bibinfo {author} {\bibfnamefont {S.~T.}\ \bibnamefont {Flammia}}, \bibinfo
  {author} {\bibfnamefont {R.}~\bibnamefont {Somma}}, \bibinfo {author}
  {\bibfnamefont {D.}~\bibnamefont {Gross}}, \bibinfo {author} {\bibfnamefont
  {S.~D.}\ \bibnamefont {Bartlett}}, \bibinfo {author} {\bibfnamefont
  {O.}~\bibnamefont {{Landon-Cardinal}}}, \bibinfo {author} {\bibfnamefont
  {D.}~\bibnamefont {Poulin}}, \ and\ \bibinfo {author} {\bibfnamefont {Y.-K.}\
  \bibnamefont {Liu}},\ }\href {\doibase 10.1038/ncomms1147} {\bibfield
  {journal} {\bibinfo  {journal} {Nat Commun}\ }\textbf {\bibinfo {volume}
  {1}},\ \bibinfo {pages} {149} (\bibinfo {year} {2010})}\BibitemShut {NoStop}%
\bibitem [{\citenamefont {Heinosaari}\ \emph {et~al.}(2013)\citenamefont
  {Heinosaari}, \citenamefont {Mazzarella},\ and\ \citenamefont
  {Wolf}}]{Heinosaari2013Quantum}%
  \BibitemOpen
  \bibfield  {author} {\bibinfo {author} {\bibfnamefont {T.}~\bibnamefont
  {Heinosaari}}, \bibinfo {author} {\bibfnamefont {L.}~\bibnamefont
  {Mazzarella}}, \ and\ \bibinfo {author} {\bibfnamefont {M.~M.}\ \bibnamefont
  {Wolf}},\ }\href {\doibase 10.1007/s00220-013-1671-8} {\bibfield  {journal}
  {\bibinfo  {journal} {Commun. Math. Phys.}\ }\textbf {\bibinfo {volume}
  {318}},\ \bibinfo {pages} {355} (\bibinfo {year} {2013})}\BibitemShut
  {NoStop}%
\bibitem [{\citenamefont {Orus}(2014)}]{Orus2014Practical}%
  \BibitemOpen
  \bibfield  {author} {\bibinfo {author} {\bibfnamefont {R.}~\bibnamefont
  {Orus}},\ }\href {\doibase 10.1016/j.aop.2014.06.013} {\bibfield  {journal}
  {\bibinfo  {journal} {Ann. Phys.}\ }\textbf {\bibinfo {volume} {349}},\
  \bibinfo {pages} {117} (\bibinfo {year} {2014})},\ \Eprint
  {http://arxiv.org/abs/1306.2164} {arxiv:1306.2164 [cond-mat, physics:hep-lat,
  physics:hep-th, physics:quant-ph]} \BibitemShut {NoStop}%
\bibitem [{\citenamefont {Schollwock}(2011)}]{Schollwock2011Densitymatrix}%
  \BibitemOpen
  \bibfield  {author} {\bibinfo {author} {\bibfnamefont {U.}~\bibnamefont
  {Schollwock}},\ }\href {\doibase 10.1098/rsta.2010.0382} {\bibfield
  {journal} {\bibinfo  {journal} {Philos. Trans. R. Soc. Math. Phys. Eng.
  Sci.}\ }\textbf {\bibinfo {volume} {369}},\ \bibinfo {pages} {2643} (\bibinfo
  {year} {2011})}\BibitemShut {NoStop}%
\bibitem [{\citenamefont {Carleo}\ and\ \citenamefont
  {Troyer}(2017)}]{Carleo2017Solving}%
  \BibitemOpen
  \bibfield  {author} {\bibinfo {author} {\bibfnamefont {G.}~\bibnamefont
  {Carleo}}\ and\ \bibinfo {author} {\bibfnamefont {M.}~\bibnamefont
  {Troyer}},\ }\href {\doibase 10.1126/science.aag2302} {\bibfield  {journal}
  {\bibinfo  {journal} {Science}\ }\textbf {\bibinfo {volume} {355}},\ \bibinfo
  {pages} {602} (\bibinfo {year} {2017})}\BibitemShut {NoStop}%
\bibitem [{\citenamefont {Deng}\ \emph
  {et~al.}(2017{\natexlab{a}})\citenamefont {Deng}, \citenamefont {Li},\ and\
  \citenamefont {Das~Sarma}}]{Deng2017Quantum}%
  \BibitemOpen
  \bibfield  {author} {\bibinfo {author} {\bibfnamefont {D.-L.}\ \bibnamefont
  {Deng}}, \bibinfo {author} {\bibfnamefont {X.}~\bibnamefont {Li}}, \ and\
  \bibinfo {author} {\bibfnamefont {S.}~\bibnamefont {Das~Sarma}},\ }\href
  {\doibase 10.1103/PhysRevX.7.021021} {\bibfield  {journal} {\bibinfo
  {journal} {Phys. Rev. X}\ }\textbf {\bibinfo {volume} {7}},\ \bibinfo {pages}
  {021021} (\bibinfo {year} {2017}{\natexlab{a}})}\BibitemShut {NoStop}%
\bibitem [{\citenamefont {Torlai}\ \emph {et~al.}(2018)\citenamefont {Torlai},
  \citenamefont {Mazzola}, \citenamefont {Carrasquilla}, \citenamefont
  {Troyer}, \citenamefont {Melko},\ and\ \citenamefont
  {Carleo}}]{Torlai2018Neuralnetwork}%
  \BibitemOpen
  \bibfield  {author} {\bibinfo {author} {\bibfnamefont {G.}~\bibnamefont
  {Torlai}}, \bibinfo {author} {\bibfnamefont {G.}~\bibnamefont {Mazzola}},
  \bibinfo {author} {\bibfnamefont {J.}~\bibnamefont {Carrasquilla}}, \bibinfo
  {author} {\bibfnamefont {M.}~\bibnamefont {Troyer}}, \bibinfo {author}
  {\bibfnamefont {R.}~\bibnamefont {Melko}}, \ and\ \bibinfo {author}
  {\bibfnamefont {G.}~\bibnamefont {Carleo}},\ }\href {\doibase
  10.1038/s41567-018-0048-5} {\bibfield  {journal} {\bibinfo  {journal} {Nature
  Phys}\ }\textbf {\bibinfo {volume} {14}},\ \bibinfo {pages} {447} (\bibinfo
  {year} {2018})}\BibitemShut {NoStop}%
\bibitem [{\citenamefont {Gao}\ and\ \citenamefont
  {Duan}(2017)}]{Gao2017Efficient}%
  \BibitemOpen
  \bibfield  {author} {\bibinfo {author} {\bibfnamefont {X.}~\bibnamefont
  {Gao}}\ and\ \bibinfo {author} {\bibfnamefont {L.-M.}\ \bibnamefont {Duan}},\
  }\href {\doibase 10.1038/s41467-017-00705-2} {\bibfield  {journal} {\bibinfo
  {journal} {Nat Commun}\ }\textbf {\bibinfo {volume} {8}},\ \bibinfo {pages}
  {662} (\bibinfo {year} {2017})}\BibitemShut {NoStop}%
\bibitem [{\citenamefont {Glasser}\ \emph {et~al.}(2018)\citenamefont
  {Glasser}, \citenamefont {Pancotti}, \citenamefont {August}, \citenamefont
  {Rodriguez},\ and\ \citenamefont {Cirac}}]{Glasser2018NeuralNetwork}%
  \BibitemOpen
  \bibfield  {author} {\bibinfo {author} {\bibfnamefont {I.}~\bibnamefont
  {Glasser}}, \bibinfo {author} {\bibfnamefont {N.}~\bibnamefont {Pancotti}},
  \bibinfo {author} {\bibfnamefont {M.}~\bibnamefont {August}}, \bibinfo
  {author} {\bibfnamefont {I.~D.}\ \bibnamefont {Rodriguez}}, \ and\ \bibinfo
  {author} {\bibfnamefont {J.~I.}\ \bibnamefont {Cirac}},\ }\href {\doibase
  10.1103/PhysRevX.8.011006} {\bibfield  {journal} {\bibinfo  {journal} {Phys.
  Rev. X}\ }\textbf {\bibinfo {volume} {8}},\ \bibinfo {pages} {011006}
  (\bibinfo {year} {2018})}\BibitemShut {NoStop}%
\bibitem [{\citenamefont {Lanyon}\ \emph {et~al.}(2017)\citenamefont {Lanyon},
  \citenamefont {Maier}, \citenamefont {Holz{\"a}pfel}, \citenamefont
  {Baumgratz}, \citenamefont {Hempel}, \citenamefont {Jurcevic}, \citenamefont
  {Dhand}, \citenamefont {Buyskikh}, \citenamefont {Daley}, \citenamefont
  {Cramer} \emph {et~al.}}]{Lanyon2017Efficient}%
  \BibitemOpen
  \bibfield  {author} {\bibinfo {author} {\bibfnamefont {B.~P.}\ \bibnamefont
  {Lanyon}}, \bibinfo {author} {\bibfnamefont {C.}~\bibnamefont {Maier}},
  \bibinfo {author} {\bibfnamefont {M.}~\bibnamefont {Holz{\"a}pfel}}, \bibinfo
  {author} {\bibfnamefont {T.}~\bibnamefont {Baumgratz}}, \bibinfo {author}
  {\bibfnamefont {C.}~\bibnamefont {Hempel}}, \bibinfo {author} {\bibfnamefont
  {P.}~\bibnamefont {Jurcevic}}, \bibinfo {author} {\bibfnamefont
  {I.}~\bibnamefont {Dhand}}, \bibinfo {author} {\bibfnamefont {A.~S.}\
  \bibnamefont {Buyskikh}}, \bibinfo {author} {\bibfnamefont {A.~J.}\
  \bibnamefont {Daley}}, \bibinfo {author} {\bibfnamefont {M.}~\bibnamefont
  {Cramer}},  \emph {et~al.},\ }\href {\doibase 10.1038/nphys4244} {\bibfield
  {journal} {\bibinfo  {journal} {Nature Phys}\ }\textbf {\bibinfo {volume}
  {13}},\ \bibinfo {pages} {1158} (\bibinfo {year} {2017})}\BibitemShut
  {NoStop}%
\bibitem [{\citenamefont {Deng}\ \emph
  {et~al.}(2017{\natexlab{b}})\citenamefont {Deng}, \citenamefont {Li},\ and\
  \citenamefont {Das~Sarma}}]{Deng2017Machine}%
  \BibitemOpen
  \bibfield  {author} {\bibinfo {author} {\bibfnamefont {D.-L.}\ \bibnamefont
  {Deng}}, \bibinfo {author} {\bibfnamefont {X.}~\bibnamefont {Li}}, \ and\
  \bibinfo {author} {\bibfnamefont {S.}~\bibnamefont {Das~Sarma}},\ }\href
  {\doibase 10.1103/PhysRevB.96.195145} {\bibfield  {journal} {\bibinfo
  {journal} {Phys. Rev. B}\ }\textbf {\bibinfo {volume} {96}} (\bibinfo {year}
  {2017}{\natexlab{b}}),\ 10.1103/PhysRevB.96.195145}\BibitemShut {NoStop}%
\bibitem [{\citenamefont {Torlai}\ and\ \citenamefont
  {Melko}(2018)}]{Torlai2018Latent}%
  \BibitemOpen
  \bibfield  {author} {\bibinfo {author} {\bibfnamefont {G.}~\bibnamefont
  {Torlai}}\ and\ \bibinfo {author} {\bibfnamefont {R.~G.}\ \bibnamefont
  {Melko}},\ }\href {\doibase 10.1103/PhysRevLett.120.240503} {\bibfield
  {journal} {\bibinfo  {journal} {Phys. Rev. Lett.}\ }\textbf {\bibinfo
  {volume} {120}},\ \bibinfo {pages} {240503} (\bibinfo {year} {2018})},\
  \Eprint {http://arxiv.org/abs/1801.09684} {arxiv:1801.09684} \BibitemShut
  {NoStop}%
\bibitem [{\citenamefont {Preskill}(2018)}]{Preskill2018Quantum}%
  \BibitemOpen
  \bibfield  {author} {\bibinfo {author} {\bibfnamefont {J.}~\bibnamefont
  {Preskill}},\ }\href {\doibase 10.22331/q-2018-08-06-79} {\bibfield
  {journal} {\bibinfo  {journal} {Quantum}\ }\textbf {\bibinfo {volume} {2}},\
  \bibinfo {pages} {79} (\bibinfo {year} {2018})}\BibitemShut {NoStop}%
\bibitem [{\citenamefont {Carrasquilla}\ \emph {et~al.}(2019)\citenamefont
  {Carrasquilla}, \citenamefont {Torlai}, \citenamefont {Melko},\ and\
  \citenamefont {Aolita}}]{Carrasquilla2019Reconstructing}%
  \BibitemOpen
  \bibfield  {author} {\bibinfo {author} {\bibfnamefont {J.}~\bibnamefont
  {Carrasquilla}}, \bibinfo {author} {\bibfnamefont {G.}~\bibnamefont
  {Torlai}}, \bibinfo {author} {\bibfnamefont {R.~G.}\ \bibnamefont {Melko}}, \
  and\ \bibinfo {author} {\bibfnamefont {L.}~\bibnamefont {Aolita}},\ }\href
  {\doibase 10.1038/s42256-019-0028-1} {\bibfield  {journal} {\bibinfo
  {journal} {Nat Mach Intell}\ }\textbf {\bibinfo {volume} {1}},\ \bibinfo
  {pages} {155} (\bibinfo {year} {2019})}\BibitemShut {NoStop}%
\bibitem [{\citenamefont {Pascanu}\ \emph {et~al.}(2013)\citenamefont
  {Pascanu}, \citenamefont {Mikolov},\ and\ \citenamefont
  {Bengio}}]{Pascanu2013Difficulty}%
  \BibitemOpen
  \bibfield  {author} {\bibinfo {author} {\bibfnamefont {R.}~\bibnamefont
  {Pascanu}}, \bibinfo {author} {\bibfnamefont {T.}~\bibnamefont {Mikolov}}, \
  and\ \bibinfo {author} {\bibfnamefont {Y.}~\bibnamefont {Bengio}},\ }in\
  \href {https://proceedings.mlr.press/v28/pascanu13.html} {\emph {\bibinfo
  {booktitle} {Proc. 30th {{Int}}. {{Conf}}. {{Mach}}. {{Learn}}.}}}\ (\bibinfo
   {publisher} {PMLR},\ \bibinfo {year} {2013})\ pp.\ \bibinfo {pages}
  {1310--1318}\BibitemShut {NoStop}%
\bibitem [{\citenamefont {Salehinejad}\ \emph {et~al.}(2018)\citenamefont
  {Salehinejad}, \citenamefont {Sankar}, \citenamefont {Barfett}, \citenamefont
  {Colak},\ and\ \citenamefont {Valaee}}]{Salehinejad2018Recent}%
  \BibitemOpen
  \bibfield  {author} {\bibinfo {author} {\bibfnamefont {H.}~\bibnamefont
  {Salehinejad}}, \bibinfo {author} {\bibfnamefont {S.}~\bibnamefont {Sankar}},
  \bibinfo {author} {\bibfnamefont {J.}~\bibnamefont {Barfett}}, \bibinfo
  {author} {\bibfnamefont {E.}~\bibnamefont {Colak}}, \ and\ \bibinfo {author}
  {\bibfnamefont {S.}~\bibnamefont {Valaee}},\ }\href {\doibase
  10.48550/arXiv.1801.01078} {\enquote {\bibinfo {title} {Recent {{Advances}}
  in {{Recurrent Neural Networks}}},}\ } (\bibinfo {year} {2018}),\ \Eprint
  {http://arxiv.org/abs/1801.01078} {arxiv:1801.01078 [cs]} \BibitemShut
  {NoStop}%
\bibitem [{\citenamefont {Fuchs}\ \emph {et~al.}(2017)\citenamefont {Fuchs},
  \citenamefont {Hoang},\ and\ \citenamefont {Stacey}}]{Fuchs2017SIC}%
  \BibitemOpen
  \bibfield  {author} {\bibinfo {author} {\bibfnamefont {C.~A.}\ \bibnamefont
  {Fuchs}}, \bibinfo {author} {\bibfnamefont {M.~C.}\ \bibnamefont {Hoang}}, \
  and\ \bibinfo {author} {\bibfnamefont {B.~C.}\ \bibnamefont {Stacey}},\
  }\href {\doibase 10.3390/axioms6030021} {\bibfield  {journal} {\bibinfo
  {journal} {Axioms}\ }\textbf {\bibinfo {volume} {6}},\ \bibinfo {pages} {21}
  (\bibinfo {year} {2017})}\BibitemShut {NoStop}%
\bibitem [{\citenamefont {Sutskever}\ \emph {et~al.}(2011)\citenamefont
  {Sutskever}, \citenamefont {Martens},\ and\ \citenamefont
  {Hinton}}]{Sutskever2011Generating}%
  \BibitemOpen
  \bibfield  {author} {\bibinfo {author} {\bibfnamefont {I.}~\bibnamefont
  {Sutskever}}, \bibinfo {author} {\bibfnamefont {J.}~\bibnamefont {Martens}},
  \ and\ \bibinfo {author} {\bibfnamefont {G.}~\bibnamefont {Hinton}},\ }in\
  \href@noop {} {\emph {\bibinfo {booktitle} {Proc. 28th {{Int}}. {{Conf}}.
  {{Int}}. {{Conf}}. {{Mach}}. {{Learn}}.}}},\ \bibinfo {series and number}
  {{{ICML}}'11}\ (\bibinfo  {publisher} {Omnipress},\ \bibinfo {address}
  {Madison, WI, USA},\ \bibinfo {year} {2011})\ pp.\ \bibinfo {pages}
  {1017--1024}\BibitemShut {NoStop}%
\bibitem [{\citenamefont {Hillery}\ \emph {et~al.}(1999)\citenamefont
  {Hillery}, \citenamefont {Bu{\v z}ek},\ and\ \citenamefont
  {Berthiaume}}]{Hillery1999Quantum}%
  \BibitemOpen
  \bibfield  {author} {\bibinfo {author} {\bibfnamefont {M.}~\bibnamefont
  {Hillery}}, \bibinfo {author} {\bibfnamefont {V.}~\bibnamefont {Bu{\v z}ek}},
  \ and\ \bibinfo {author} {\bibfnamefont {A.}~\bibnamefont {Berthiaume}},\
  }\href {\doibase 10.1103/PhysRevA.59.1829} {\bibfield  {journal} {\bibinfo
  {journal} {Phys. Rev. A}\ }\textbf {\bibinfo {volume} {59}},\ \bibinfo
  {pages} {1829} (\bibinfo {year} {1999})}\BibitemShut {NoStop}%
\bibitem [{\citenamefont {Degen}\ \emph {et~al.}(2017)\citenamefont {Degen},
  \citenamefont {Reinhard},\ and\ \citenamefont
  {Cappellaro}}]{Degen2017Quantum}%
  \BibitemOpen
  \bibfield  {author} {\bibinfo {author} {\bibfnamefont {C.~L.}\ \bibnamefont
  {Degen}}, \bibinfo {author} {\bibfnamefont {F.}~\bibnamefont {Reinhard}}, \
  and\ \bibinfo {author} {\bibfnamefont {P.}~\bibnamefont {Cappellaro}},\
  }\href {\doibase 10.1103/RevModPhys.89.035002} {\bibfield  {journal}
  {\bibinfo  {journal} {Rev. Mod. Phys.}\ }\textbf {\bibinfo {volume} {89}},\
  \bibinfo {pages} {035002} (\bibinfo {year} {2017})}\BibitemShut {NoStop}%
\bibitem [{\citenamefont {Scott}(2006)}]{Scott2006Tight}%
  \BibitemOpen
  \bibfield  {author} {\bibinfo {author} {\bibfnamefont {A.~J.}\ \bibnamefont
  {Scott}},\ }\href {\doibase 10.1088/0305-4470/39/43/009} {\bibfield
  {journal} {\bibinfo  {journal} {J. Phys. A: Math. Gen.}\ }\textbf {\bibinfo
  {volume} {39}},\ \bibinfo {pages} {13507} (\bibinfo {year}
  {2006})}\BibitemShut {NoStop}%
\bibitem [{\citenamefont {D'Ariano}\ \emph {et~al.}(2004)\citenamefont
  {D'Ariano}, \citenamefont {Perinotti},\ and\ \citenamefont
  {Sacchi}}]{DAriano2004Informationally}%
  \BibitemOpen
  \bibfield  {author} {\bibinfo {author} {\bibfnamefont {G.~M.}\ \bibnamefont
  {D'Ariano}}, \bibinfo {author} {\bibfnamefont {P.}~\bibnamefont {Perinotti}},
  \ and\ \bibinfo {author} {\bibfnamefont {M.~F.}\ \bibnamefont {Sacchi}},\
  }\href {\doibase 10.1088/1464-4266/6/6/005} {\bibfield  {journal} {\bibinfo
  {journal} {J. Opt. B: Quantum Semiclass. Opt.}\ }\textbf {\bibinfo {volume}
  {6}},\ \bibinfo {pages} {S487} (\bibinfo {year} {2004})}\BibitemShut
  {NoStop}%
\bibitem [{\citenamefont {Barends}\ \emph {et~al.}(2013)\citenamefont
  {Barends}, \citenamefont {Kelly}, \citenamefont {Megrant}, \citenamefont
  {Sank}, \citenamefont {Jeffrey}, \citenamefont {Chen}, \citenamefont {Yin},
  \citenamefont {Chiaro}, \citenamefont {Mutus}, \citenamefont {Neill} \emph
  {et~al.}}]{Barends2013Coherent}%
  \BibitemOpen
  \bibfield  {author} {\bibinfo {author} {\bibfnamefont {R.}~\bibnamefont
  {Barends}}, \bibinfo {author} {\bibfnamefont {J.}~\bibnamefont {Kelly}},
  \bibinfo {author} {\bibfnamefont {A.}~\bibnamefont {Megrant}}, \bibinfo
  {author} {\bibfnamefont {D.}~\bibnamefont {Sank}}, \bibinfo {author}
  {\bibfnamefont {E.}~\bibnamefont {Jeffrey}}, \bibinfo {author} {\bibfnamefont
  {Y.}~\bibnamefont {Chen}}, \bibinfo {author} {\bibfnamefont {Y.}~\bibnamefont
  {Yin}}, \bibinfo {author} {\bibfnamefont {B.}~\bibnamefont {Chiaro}},
  \bibinfo {author} {\bibfnamefont {J.}~\bibnamefont {Mutus}}, \bibinfo
  {author} {\bibfnamefont {C.}~\bibnamefont {Neill}},  \emph {et~al.},\ }\href
  {\doibase 10.1103/PhysRevLett.111.080502} {\bibfield  {journal} {\bibinfo
  {journal} {Phys. Rev. Lett.}\ }\textbf {\bibinfo {volume} {111}},\ \bibinfo
  {pages} {080502} (\bibinfo {year} {2013})}\BibitemShut {NoStop}%
\bibitem [{\citenamefont {Barends}\ \emph {et~al.}(2014)\citenamefont
  {Barends}, \citenamefont {Kelly}, \citenamefont {Megrant}, \citenamefont
  {Veitia}, \citenamefont {Sank}, \citenamefont {Jeffrey}, \citenamefont
  {White}, \citenamefont {Mutus}, \citenamefont {Fowler}, \citenamefont
  {Campbell} \emph {et~al.}}]{Barends2014Superconducting}%
  \BibitemOpen
  \bibfield  {author} {\bibinfo {author} {\bibfnamefont {R.}~\bibnamefont
  {Barends}}, \bibinfo {author} {\bibfnamefont {J.}~\bibnamefont {Kelly}},
  \bibinfo {author} {\bibfnamefont {A.}~\bibnamefont {Megrant}}, \bibinfo
  {author} {\bibfnamefont {A.}~\bibnamefont {Veitia}}, \bibinfo {author}
  {\bibfnamefont {D.}~\bibnamefont {Sank}}, \bibinfo {author} {\bibfnamefont
  {E.}~\bibnamefont {Jeffrey}}, \bibinfo {author} {\bibfnamefont {T.~C.}\
  \bibnamefont {White}}, \bibinfo {author} {\bibfnamefont {J.}~\bibnamefont
  {Mutus}}, \bibinfo {author} {\bibfnamefont {A.~G.}\ \bibnamefont {Fowler}},
  \bibinfo {author} {\bibfnamefont {B.}~\bibnamefont {Campbell}},  \emph
  {et~al.},\ }\href {\doibase 10.1038/nature13171} {\bibfield  {journal}
  {\bibinfo  {journal} {Nature}\ }\textbf {\bibinfo {volume} {508}},\ \bibinfo
  {pages} {500} (\bibinfo {year} {2014})}\BibitemShut {NoStop}%
\bibitem [{\citenamefont {Schuster}\ \emph {et~al.}(2007)\citenamefont
  {Schuster}, \citenamefont {Houck}, \citenamefont {Schreier}, \citenamefont
  {Wallraff}, \citenamefont {Gambetta}, \citenamefont {Blais}, \citenamefont
  {Frunzio}, \citenamefont {Majer}, \citenamefont {Johnson}, \citenamefont
  {Devoret} \emph {et~al.}}]{Schuster2007Resolving}%
  \BibitemOpen
  \bibfield  {author} {\bibinfo {author} {\bibfnamefont {D.~I.}\ \bibnamefont
  {Schuster}}, \bibinfo {author} {\bibfnamefont {A.~A.}\ \bibnamefont {Houck}},
  \bibinfo {author} {\bibfnamefont {J.~A.}\ \bibnamefont {Schreier}}, \bibinfo
  {author} {\bibfnamefont {A.}~\bibnamefont {Wallraff}}, \bibinfo {author}
  {\bibfnamefont {J.~M.}\ \bibnamefont {Gambetta}}, \bibinfo {author}
  {\bibfnamefont {A.}~\bibnamefont {Blais}}, \bibinfo {author} {\bibfnamefont
  {L.}~\bibnamefont {Frunzio}}, \bibinfo {author} {\bibfnamefont
  {J.}~\bibnamefont {Majer}}, \bibinfo {author} {\bibfnamefont
  {B.}~\bibnamefont {Johnson}}, \bibinfo {author} {\bibfnamefont {M.~H.}\
  \bibnamefont {Devoret}},  \emph {et~al.},\ }\href {\doibase
  10.1038/nature05461} {\bibfield  {journal} {\bibinfo  {journal} {Nature}\
  }\textbf {\bibinfo {volume} {445}},\ \bibinfo {pages} {515} (\bibinfo {year}
  {2007})}\BibitemShut {NoStop}%
\bibitem [{\citenamefont {Kamal}\ \emph {et~al.}(2009)\citenamefont {Kamal},
  \citenamefont {Marblestone},\ and\ \citenamefont
  {Devoret}}]{Kamal2009Signaltopump}%
  \BibitemOpen
  \bibfield  {author} {\bibinfo {author} {\bibfnamefont {A.}~\bibnamefont
  {Kamal}}, \bibinfo {author} {\bibfnamefont {A.}~\bibnamefont {Marblestone}},
  \ and\ \bibinfo {author} {\bibfnamefont {M.}~\bibnamefont {Devoret}},\ }\href
  {\doibase 10.1103/PhysRevB.79.184301} {\bibfield  {journal} {\bibinfo
  {journal} {Phys. Rev. B}\ }\textbf {\bibinfo {volume} {79}},\ \bibinfo
  {pages} {184301} (\bibinfo {year} {2009})}\BibitemShut {NoStop}%
\bibitem [{\citenamefont {Vijay}\ \emph {et~al.}(2009)\citenamefont {Vijay},
  \citenamefont {Devoret},\ and\ \citenamefont {Siddiqi}}]{Vijay2009Invited}%
  \BibitemOpen
  \bibfield  {author} {\bibinfo {author} {\bibfnamefont {R.}~\bibnamefont
  {Vijay}}, \bibinfo {author} {\bibfnamefont {M.~H.}\ \bibnamefont {Devoret}},
  \ and\ \bibinfo {author} {\bibfnamefont {I.}~\bibnamefont {Siddiqi}},\ }\href
  {\doibase 10.1063/1.3224703} {\bibfield  {journal} {\bibinfo  {journal}
  {Review of Scientific Instruments}\ }\textbf {\bibinfo {volume} {80}},\
  \bibinfo {pages} {111101} (\bibinfo {year} {2009})}\BibitemShut {NoStop}%
\bibitem [{\citenamefont {Roy}\ \emph {et~al.}(2015)\citenamefont {Roy},
  \citenamefont {Kundu}, \citenamefont {Chand}, \citenamefont {Vadiraj},
  \citenamefont {Ranadive}, \citenamefont {Nehra}, \citenamefont {Patankar},
  \citenamefont {Aumentado}, \citenamefont {Clerk},\ and\ \citenamefont
  {Vijay}}]{Roy2015Broadband}%
  \BibitemOpen
  \bibfield  {author} {\bibinfo {author} {\bibfnamefont {T.}~\bibnamefont
  {Roy}}, \bibinfo {author} {\bibfnamefont {S.}~\bibnamefont {Kundu}}, \bibinfo
  {author} {\bibfnamefont {M.}~\bibnamefont {Chand}}, \bibinfo {author}
  {\bibfnamefont {A.~M.}\ \bibnamefont {Vadiraj}}, \bibinfo {author}
  {\bibfnamefont {A.}~\bibnamefont {Ranadive}}, \bibinfo {author}
  {\bibfnamefont {N.}~\bibnamefont {Nehra}}, \bibinfo {author} {\bibfnamefont
  {M.~P.}\ \bibnamefont {Patankar}}, \bibinfo {author} {\bibfnamefont
  {J.}~\bibnamefont {Aumentado}}, \bibinfo {author} {\bibfnamefont {A.~A.}\
  \bibnamefont {Clerk}}, \ and\ \bibinfo {author} {\bibfnamefont
  {R.}~\bibnamefont {Vijay}},\ }\href {\doibase 10.1063/1.4939148} {\bibfield
  {journal} {\bibinfo  {journal} {Applied Physics Letters}\ }\textbf {\bibinfo
  {volume} {107}},\ \bibinfo {pages} {262601} (\bibinfo {year}
  {2015})}\BibitemShut {NoStop}%
\bibitem [{\citenamefont {Li}\ \emph {et~al.}(2018)\citenamefont {Li},
  \citenamefont {Ma}, \citenamefont {Han}, \citenamefont {Chen}, \citenamefont
  {Xu}, \citenamefont {Cai}, \citenamefont {Wang}, \citenamefont {Song},
  \citenamefont {Xue}, \citenamefont {Yin} \emph {et~al.}}]{Li2018Perfect}%
  \BibitemOpen
  \bibfield  {author} {\bibinfo {author} {\bibfnamefont {X.}~\bibnamefont
  {Li}}, \bibinfo {author} {\bibfnamefont {Y.}~\bibnamefont {Ma}}, \bibinfo
  {author} {\bibfnamefont {J.}~\bibnamefont {Han}}, \bibinfo {author}
  {\bibfnamefont {T.}~\bibnamefont {Chen}}, \bibinfo {author} {\bibfnamefont
  {Y.}~\bibnamefont {Xu}}, \bibinfo {author} {\bibfnamefont {W.}~\bibnamefont
  {Cai}}, \bibinfo {author} {\bibfnamefont {H.}~\bibnamefont {Wang}}, \bibinfo
  {author} {\bibfnamefont {Y.}~\bibnamefont {Song}}, \bibinfo {author}
  {\bibfnamefont {Z.-Y.}\ \bibnamefont {Xue}}, \bibinfo {author} {\bibfnamefont
  {Z.-q.}\ \bibnamefont {Yin}},  \emph {et~al.},\ }\href {\doibase
  10.1103/PhysRevApplied.10.054009} {\bibfield  {journal} {\bibinfo  {journal}
  {Phys. Rev. Appl.}\ }\textbf {\bibinfo {volume} {10}},\ \bibinfo {pages}
  {054009} (\bibinfo {year} {2018})}\BibitemShut {NoStop}%
\bibitem [{sup()}]{supplement}%
  \BibitemOpen
  \href@noop {} {\enquote {\bibinfo {title} {Supplementary {M}aterials},}\
  }\BibitemShut {NoStop}%
\bibitem [{\citenamefont {Diamond}\ and\ \citenamefont
  {Boyd}(2016)}]{Diamond2016CVXPY}%
  \BibitemOpen
  \bibfield  {author} {\bibinfo {author} {\bibfnamefont {S.}~\bibnamefont
  {Diamond}}\ and\ \bibinfo {author} {\bibfnamefont {S.}~\bibnamefont {Boyd}},\
  }\href@noop {} {\bibfield  {journal} {\bibinfo  {journal} {J. Mach. Learn.
  Res.}\ }\textbf {\bibinfo {volume} {17}},\ \bibinfo {pages} {2909} (\bibinfo
  {year} {2016})}\BibitemShut {NoStop}%
\bibitem [{\citenamefont {Martinis}\ and\ \citenamefont
  {Geller}(2014)}]{martinis2014fast}%
  \BibitemOpen
  \bibfield  {author} {\bibinfo {author} {\bibfnamefont {J.~M.}\ \bibnamefont
  {Martinis}}\ and\ \bibinfo {author} {\bibfnamefont {M.~R.}\ \bibnamefont
  {Geller}},\ }\href {\doibase 10.1103/PhysRevA.90.022307} {\bibfield
  {journal} {\bibinfo  {journal} {Phys. Rev. A}\ }\textbf {\bibinfo {volume}
  {90}},\ \bibinfo {pages} {022307} (\bibinfo {year} {2014})}\BibitemShut
  {NoStop}%
\end{thebibliography}

\end{document}